\documentclass[]{interact}

\usepackage{graphicx}
\usepackage{amsmath,bm}
\usepackage[utf8x]{inputenc} 
\usepackage{float} 
\usepackage{datetime}
\usepackage{cite} 

\usepackage[numbers,sort&compress]{natbib}
\bibpunct[, ]{[}{]}{,}{n}{,}{,}
\renewcommand\bibfont{\fontsize{10}{12}\selectfont}
\makeatletter
\def\NAT@def@citea{\def\@citea{\NAT@separator}}
\makeatother

\usepackage{color,verbatim,epstopdf}
\usepackage{amssymb,amsthm}

\newcommand{\be}{\begin{equation}}
\newcommand{\ee}{\end{equation}}

\newcommand{\lb}[1]{{\color{black}{#1}}}
\newcommand{\obs}[1]{{\color{black}{#1}}}

\newcommand{\mb}[1]{{\color{black}{#1}}}

\def\druvec{{\delta} \mathbf{v}}
\def\dru{{\delta} v}
\def\ux { \mathbf{x} }
\def\ur { \mathbf{r} }
\def\urh { \hat{\mathbf{r}} }

\def\la {\langle}
\def\ra {\rangle}
\def\p{\partial}

\def\eps{\varepsilon}

\newcommand{\RL}{{\rm Re_\ell}}

\newcommand{\tbv}{\tilde{\boldsymbol{v}}}
\newcommand{\ov}{\overline{ v}}
\newcommand{\tv}{\tilde{ v}}
\newcommand{\obv}{\overline{\boldsymbol{v}}}

\newcommand{\oP}{\overline{{ p}}}
\newcommand{\tP}{\tilde{{ p}}}
\newcommand{\oPi}{\overline{{\Pi}}^\Delta}
\newcommand{\tPi}{\tilde{{\Pi}}^\Delta}
\newcommand{\opi}{\overline{{\pi}}^\Delta}
\newcommand{\Pileo}{{\overline{\Pi}}^\Delta_{L}}

\newcommand{\otau}{\overline{{\tau}}^\Delta}
\newcommand{\tauleo}{{\tau}_{ij}^{\Delta,L}}

\newcommand{\ttau}{\tilde{{\tau}}^\Delta}
\newcommand{\oS}{\overline{s}}
\newcommand{\tS}{\tilde{s}}
\newcommand{\bk}{\boldsymbol{k}}
\newcommand{\bx}{\boldsymbol{x}}
\newcommand{\by}{\boldsymbol{y}}
\newcommand{\br}{\boldsymbol{r}}
\newcommand{\bu}{\boldsymbol{u}}
\newcommand{\bv}{\boldsymbol{v}}

\newcommand{\bF}{\boldsymbol{f}}

\newcommand{\bp}{\boldsymbol{p}}

\makeatletter
\def\extremetilde#1{\mathop{\vbox{\m@th\ialign{##\crcr\noalign{\kern3\p@}%
      \extremetildefill\crcr\noalign{\kern3\p@\nointerlineskip}%
      $\hfil\displaystyle{#1}\hfil$\crcr}}}\limits}

\def\extremetildefill{$\m@th \setbox\z@\hbox{$\braceld$}%
  \braceld\leaders\vrule \@height\ht\z@ \@depth\z@\hfill\braceru$}

\makeatother

\begin{document}

\title{\obs{
Effect of filter type on the statistics of energy transfer between resolved and subfilter scales from
a-priori analysis of direct numerical simulations of isotropic turbulence \footnote{Postprint version of the manuscript published in J. Turbul. {\bf 19} No.~2, 167-197 (2018)}}
}
\author{
\name{M. Buzzicotti$^1$, M. Linkmann$^1$, H. Aluie$^2$, L. Biferale$^1$, J. Brasseur$^3$ and C. Meneveau$^4$}
\affil{$^1$Dept. of Physics and INFN, University of Rome Tor Vergata, Rome, Italy \\
$^2$Dept. of Mechanical Engineering, University of Rochester, USA\\
$^3$Aerospace Engineering Sciences, University of Colorado, Boulder, USA \\
$^4$Dept. of Mechanical Engineering, The Johns Hopkins University, Baltimore, USA
}
}

\maketitle

\begin{abstract} 
\noindent
The effects of different filtering strategies on the statistical properties of
the resolved-to-subfilter scale (SFS) energy transfer are analysed in forced
homogeneous and isotropic turbulence.  We carry out {\em a priori} analyses of
the statistical characteristics of SFS energy transfer by filtering data obtained
from  direct numerical simulations with up to $2048^3$ grid points as a
function of the filter cut-off scale.  In order to quantify the dependence of
extreme events and anomalous scaling on the filter, we compare a sharp Fourier
Galerkin projector, a Gaussian filter and a novel class of Galerkin projectors
with non-sharp spectral filter profiles.  Of interest is the importance of
Galilean invariance and we confirm  that local SFS energy transfer displays
intermittency scaling in both skewness and flatness as a function of the cut-off
scale. Furthermore, we quantify the robustness of scaling as a function of the
filtering type. 
\end{abstract}

\begin{keywords}
isotropic turbulence, large eddy simulation
\end{keywords}

\section{Introduction} 
\noindent
Understanding  \obs{and predicting} multiscale turbulent statistics \obs{are}  key challenges
for many  modern applied and fundamental problems in fluid \obs{turbulence} dynamics.  
{Of major interest} is the existence of
intermittency \cite{Frisch95,Pope00,Lesieur08}, 
{i.e. the development of anomalously intense fluctuations that depart more and more from Gaussianity
 by going to smaller and smaller scales \cite{Frisch95}.}
Similarly, the statistics of velocity  gradients become  \obs{increasingly} intermittent  
{by} augmenting the turbulent
intensity, generally expressed by the Reynolds number, a measure of the
relative importance of non-linear and linear terms in the three dimensional
Navier-Stokes equations.  Anomalous multiscale fluctuations are generic in 
three-dimensional (3D) turbulence, being observed at small scales 
in homogeneous and \lb{inhomogeneous flows such as} wall-bounded flows
\cite{Arneodo_et_al1996,Qian02,Gotoh02,Benzi99,antonia2000,desilva2015} and also in Lagrangian statistics 
\cite{laporta2001,mordant2001,Biferale04,Arneodo08,Toschi09,Benzi10}.  
The Reynolds numbers attainable in direct numerical simulations (DNS) are still far below those
occurring in nature and in most engineering applications. Hence, modelling is
often unavoidable, and one of the commonly used  approaches is based
on  large eddy simulation (LES) \cite{Smagorinsky63,MeneveauARFM,Sagaut08,Lesieur08,Pope00}.\\
The basic idea of LES 
is to advance the turbulence dynamics on a coarse-grained grid with resolution
sufficient to capture a large percent of the turbulent kinetic energy (and
variances in other key fluctuating variables). Thus, models are required to
capture dominant effects of the subfilter-scale (SFS) motions on resolved
large-scale dynamics.
In this paper, we focus primarily on the
impact of the details of the filter type on the statistics of the resolved-scale, both at the energy-containing range and close to the filter scale.
We do this using only physical-space data from fully resolved DNS by analysing physical space SFS statistics without
introducing any modelling (\emph{a-priori} analysis \cite{Piomelli88}).
\\
\noindent 
LES \obs{aims to predict} 
integral-scale \obs{variables, where kinetic energy is concentrated,} 
to \obs{acceptable} 
degree\obs{s} of accuracy. 
In practice, the accuracy of \obs{the SFS} 
model is \obs{generally} 
assessed in terms of its ability to achieve good agreement with empirical
measurements \obs{of} 
one-point or two-point  statistic\obs{s} 
such as 
mean profile{s}, energy spectr\obs{a} and the Reynolds stress tensor.  On the
other hand, it is well known that 
LES 
{has special difficulties near} solid boundaries, where 
{key integral length scales are proportional to the distance from the wall 
and integral-scale motions tend to become under-resolved, and where}
small-scale energy and vorticity
injections/ejections {directly impact} 
the mean flow 
\cite{Pope00,Zhou01,Brasseur10,meneveau2013,desilva2015,Hellstrom16,Yang16}.
The most common LES models 
{replace the SFS stress tensor with an eddy viscosity form that replaces
the true inertial resolved-SFS dynamics with a form that is
diffusive in momentum and dissipative in energy.}
{H}ence, {the} inertial (time reversible) contribution to the resolved-to-SFS 
dynamics is
everywhere approximated as a dissipative loss \cite{Smagorinsky63}.
{I}t is  
{increasingly} common {for}  
LES 
{to be applied at relatively high resolution with the filter scale}
well {within} 
the inertial {sub}range \cite{Stevens14,MartinezTossas16}. 
In such cases, {there is a} need to go beyond first- and
second-order statistics to validate/benchmark the {LES} accuracy 
{in relationship} to the presence of strong non-Gaussian fluctuations 
{and of resolved-to-SFS interactions at the smallest resolved scales \cite{Meneveau94},
  which might even lead to backscatter
events \cite{Waleffe92,Casciola03,Fang12,Biferale12,Chen03a,Chen03b}}.
{Moreover,} 
the need to apply LES to study
Lagrangian evolution of small  particles or to the advection/reaction
of Eulerian fields (combustion, multicomponent flows, etc.), calls for 
{refined} 
control of the impact of {the modelled SFS stress tensor} 
on {the} multiscale statistical properties {of the predicted}
resolved velocity field.  
{Furthermore, in} 
many important turbulent {flows}, 
a global backward cascade with a mean negative energy transfer in the \obs{domain exists}. 
\obs{This is the case with} fast rotating flows \cite{Smith96,Mininni09a,BiferalePRX2016}
\obs{and} 
shallow fluid layers \cite{Nastrom84,Lautenschlager98,Smith99,Celani10,Xia11}. 
\obs{I}n certain circumstances of conducting flows \cite{Frisch75,Pouquet76},
the backward-{\em cascading} quantity is the magnetic helicity, which results in 
some energy also being {\em transferred} from small scales to large scales 
if the magnetic helicity is nonzero. 
\\
\noindent \obs{The first goal of this paper is to present a systematic
investigation of the key statistical properties  of SFS energy transfer that the modelled SFS stress tensor
should reproduce in order to capture intense non-Gaussian SFS
fluctuations, including those 
responsible for back-scatter. We do
that by performing an {\em a priori} study of the multiscale properties of \obs{SFS} 
energy transfer from high-resolution DNS on up to $2048$ collocation points per spatial direction.}
In particular, we \obs{aim} 
to define a set of benchmarks 
\obs{for future high-resolution} LES 
\obs{of high Reynolds number turbulence where an inertial subrange of turbulence 
scales exists and is well resolved, so that strong intermittency close to the SFS cut-off creates large}
departure\obs{s} from \obs{Gaussian} statistics. \\
\noindent
In \obs{LES where intermittency at the smallest resolved scales is of interest},
benchmarking \obs{the} 
LES model 
\obs{will require evaluation of multi-point statistics of 
order higher than second  (i.e., beyond (co)variances, 
spectral properties, correlation functions, etc.)}.
\obs{In p}revious work \cite{Cerutti98}, 
intermittency and non-Gaussian properties of the SFS
energy transfer \obs{were analysed} at moderate resolution and by using extended self
similarity (ESS) \cite{Benzi93,Benzi95}, showing that \obs{SFS energy transfer statistics are} 
affected by non-trivial anomalous deviations from the \obs{classical} 
scaling as a function of the cut-off scale. Here, we follow the same \obs{approach} 
but {we focus on the impact of the filter and} extend 
{the analysis} to much higher Reynolds numbers and by changing the filter properties (see
as follows).  \\ 
Another motivation \obs{for the current analysis} is based on more fundamental aspects. 
It is usually thought that the inertial spectral properties in fully developed homogeneous turbulence are asymptotically independent of the way energy is absorbed at high wavenumbers, i.e. if the Reynolds number is large enough, the inertial-range statistics of second-order velocity correlation functions are independent of the mechanism by which energy is transferred and absorbed at the small scales since the interactions distributing the kinetic energy are mostly local in scale \cite{Eyink05,aluie2009I,aluie2009II}.

This is the main motivation behind the introduction of hyperviscosity in many numerical studies \cite{Borue95}.
However, in LES, the statistics at the smallest resolved scales can 
certainly depend on the details of the SFS model. Since higher-order moments  have a non-local support in Fourier space, one would expect that they might  become progressively more sensitive to
the details of the model for the SFS stress tensor.
\mb{A natural question then arises: is it possible to devise a LES scheme which {\it minimises} the cut-off 
effects on the resolved inertial range,} achieving 
a scaling as extended as possible for high-order correlation functions too?
{Improved closures that also predict intermittency}
would be helpful also to LES practitioners
{with a} 
need to push the \obs{SFS} cut-off to scales small enough where
intermittency \obs{effects are important}. 
\\
\noindent Being interested in intense-but-rare statistical properties, we need
first to define a set of \obs{SFS} observables which 
are statistically robust and
not {strongly} 
affected by filter-induced effects and/or fluctuations
induced by coupling among the resolved scales. 
{To this end}, 
we {apply} 
a filter  formalism \obs{that} isolates the terms that genuinely couple resolved and \obs{unresolved} 
scales from those that are affected by other  contributions due to
self-coupling of the resolved fields. Moreover, we discuss in detail the
importance of focusing on Galilean invariant quantities, in order to avoid
strong contamination from unphysical fluctuations affecting  the very intense
events (and not the mean single-point properties) 
\cite{aluie2009I,aluie2009II,Speziale85}. \\
In what follows, we  carry out an {\em a priori} analysis of the different
components of the SFS energy transfer by filtering DNS data 
{with} 
different filter thresholds.  
Since LES results
depend not only on the details of the SFS-model but also on the choice of the
filter, and since the filter-induced fluctuations will vary, 
we \obs{are particularly interested in sensitivity to} 
different filtering \obs{types and} procedures.   
Beside the
standard  sharp Fourier Galerkin projector and a convolution with a Gaussian
kernel, 
we also devise a novel class of Galerkin filters with
a non-sharp probabilistic profile in Fourier space. The \obs{set of} new projectors 
offer multiscale filtering in Fourier space while maintaining formal and
practical advantages specific to projector filters. 
\\
\noindent 
The analysis is structured as follows.  We begin \obs{in section
\ref{sec:data_descr}} with a brief description of the DNS database and the
numerical methods \obs{applied} to generate and analyse the data.  In
sec.~\ref{sec:theory}, we motivate and introduce the  filtering formulation and
investigate the properties of different filter-dependent components of SFS
energy transfer under Galilean transformations.  We show that an apparent
breaking of Galilean invariance of the SFS-energy transfer term can be remedied
by introducing  a more refined distinction between the contributions to the SFS
energy transfer which require modelling and those which do not. The first
results from the {\em a-priori} analysis are presented in
sec.~\ref{sec:intermittency}, where we measure the fluctuations of different
components of the SFS energy transfer. We  provide measurements of intermittent
scaling of the SFS energy transfer at high Reynolds numbers without using ESS.
Section \ref{sec:filters} is dedicated to an analysis of different filtering
procedures on the statistics of SFS energy transfer and we  introduce a
novel class of Galerkin filters with a non-sharp profile in Fourier space.  We
summarise our results in sec.~\ref{sec:conclusions}.

\section{Description of the data-sets} \label{sec:data_descr}

\begin{table}[H]
 \begin{center}
\begin{tabular}{ccccccccc}
   Data id & N & $\RL$ & $\varepsilon$ & $U$ & $\ell$ & $\nu$ & $\alpha$ & $T_0/T_{\rm eddy}$ \\
  \hline
V1 & 1024 & 2570 & 1.9 & 1.8 & 1.2 & 0.0008 & 1 & 25 \\
V2 & 2048 & 7000 & 1.4 & 1.5 & 1.2 & 0.0003 & 1 & 9 \\
H1 & 1024 & 8000 & 1.9 & 1.9 & 1.3 & $2 \times 10^{-8}$ & 2 & 7 \\
H2 & 2048 & 26000 & 1.5 & 1.6 & 1.1 & $5.7 \times 10^{-20}$ & 4 & 6 \\
  \hline
  \end{tabular}
  \end{center}
 \caption{
 The identifiers V and H distinguish between
  hyper and normal viscosity, where $\alpha$ is the order of the Laplacian. 
 $N$ denotes the number of grid points in each Cartesian coordinate, 
$U$ the rms velocity, $\ell=(\pi/2U^2)\int dk \ E(k)/k$ the integral scale,
$\nu$ the kinematic viscosity, $\varepsilon$ the dissipation rate and $T_0/T_{\rm eddy}$ the
steady-state run time in units of large-eddy turnover time $T_{\rm eddy}=\ell/U$. The
values given for $\varepsilon$, $U$, $\ell$ are time averages.  The \obs{integral-scale} Reynolds
number is $\RL = U\ell/\nu$ for data-set V while for the hyperviscous
simulations it is defined as $\RL = C (\ell/l_d)^{4/3}$, where $C$ is a
constant estimated by comparing the two definitions for data-set V and $l_d$ is
the scale corresponding to the maximum of $k^2 E(k)$, see
Figure~\ref{fig:simulations_flux_spectra} (b).  }
 \label{tbl:simulations}
 \end{table}
In order to generate the data-sets for the {\em a priori} analysis, 
we numerically solved the 3D Navier-Stokes equations using both normal and hyperviscosity 
\begin{align}
\label{eq:nse}
&\partial_t \bv = - \nabla \cdot (\bv \otimes \bv) - \nabla p + \nu (-1)^{\alpha + 1}\Delta^{\alpha}\bv + \bF \ , \\
\label{eq:incomp}
&\nabla \cdot \bv = 0 \ ,
\end{align}
where $\bv$ denotes the velocity field, $p$ is the pressure divided by the constant density, $\nu$ the
\obs {kinematic} viscosity, $\bF$ an external force and $\alpha$  the power of the Laplacian.
\obs{As indicated in Table \ref{tbl:simulations}, d}ata with normal viscosity, $\alpha =1$, are denoted as (V), data with hyperviscous dissipation  $\alpha =2$ and $\alpha =4$ are 
 identified
through the labels H1 and H2, respectively. 
We use a pseudospectral code on up to $2048^3$ collocation points in a triply
periodic domain $\Omega$ of size $L=2\pi$. Full dealiasing is implemented by application of
 the two-thirds rule \cite{Patterson71}. 
The homogeneous and isotropic external force $\bF$ 
is defined via a second-order Ornstein-Uhlenbeck process in a band of Fourier modes $k \in [0.5,1.5]$  \cite{Sawford91,BiferalePRX2016}. 
\obs{The resolution of the simulations quantified 
in terms of the grid spacing $dx$ and the  
Kolmogorov microscale $\eta_{\alpha} = (\nu^3/\eps)^{1/6\alpha -2}$ 
\cite{Borue95}, where $\eps$ is the dissipation rate, 
is $\eta_{\alpha}/dx \simeq 0.7$ for all simulations.} \\

\begin{figure}[H]
\hspace{-0.5cm}
  \includegraphics[scale = 0.47]{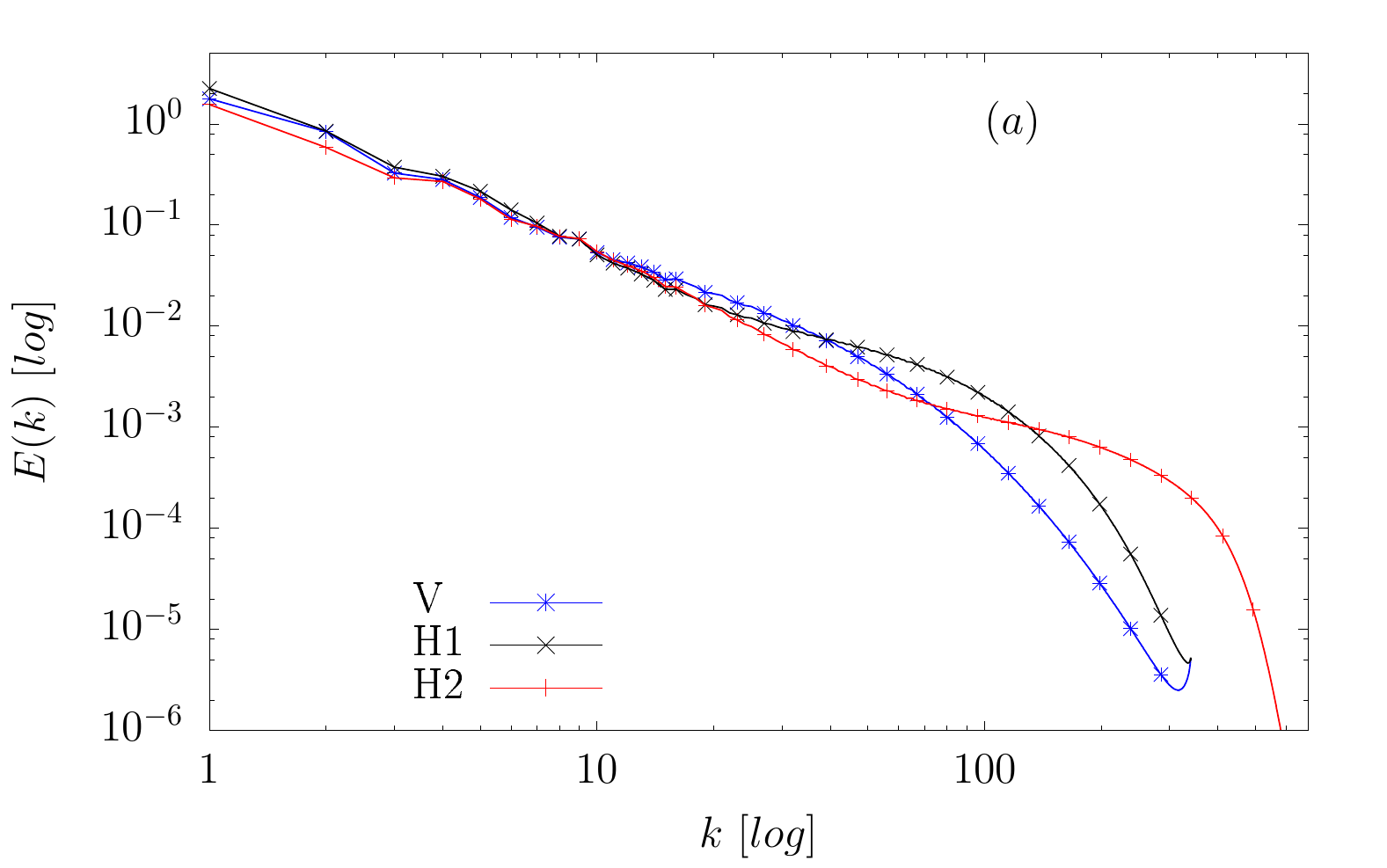}
  \includegraphics[scale = 0.47]{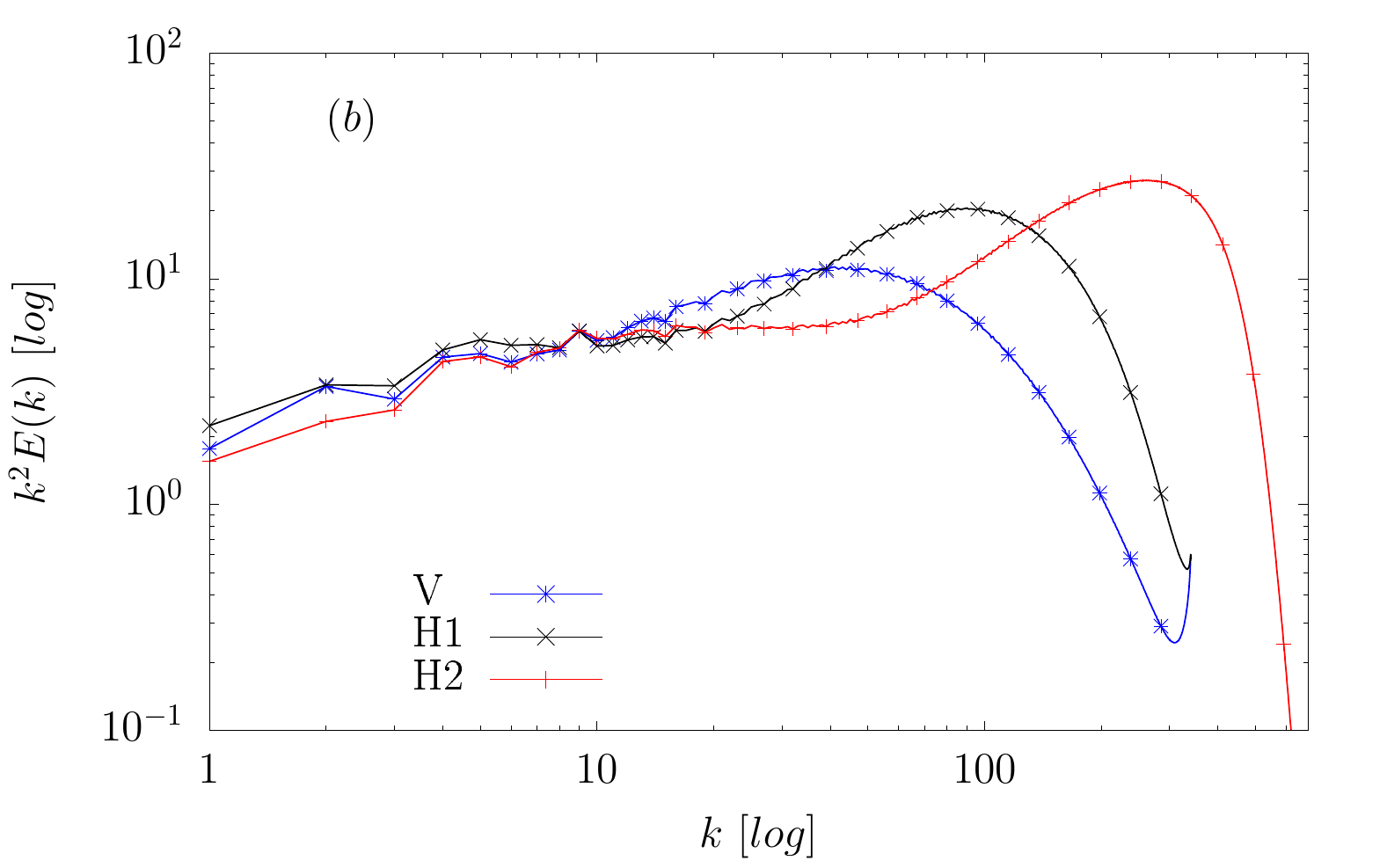}
  \caption{
\mb{Kinetic energy spectra $ E(k)$ (a) and dissipation spectra $k^{2} E(k)$ (b) for data-sets V1, V2, H1 and H2. 
The dissipation spectra $k^2E(k)$ instead of their hyperviscous counterparts 
$k^{2\alpha}E(k)$ are shown because the former are connected to the (physical) velocity-field gradients, 
which are used to estimate a Reynolds number for the hyperviscous simulations
in a consistent way compared to run V1 and V2. See also table \ref{tbl:simulations}}. 
}
\label{fig:simulations_flux_spectra}
\end{figure}

\begin{figure}[H]
\begin{center}
  \includegraphics[scale = 0.8]{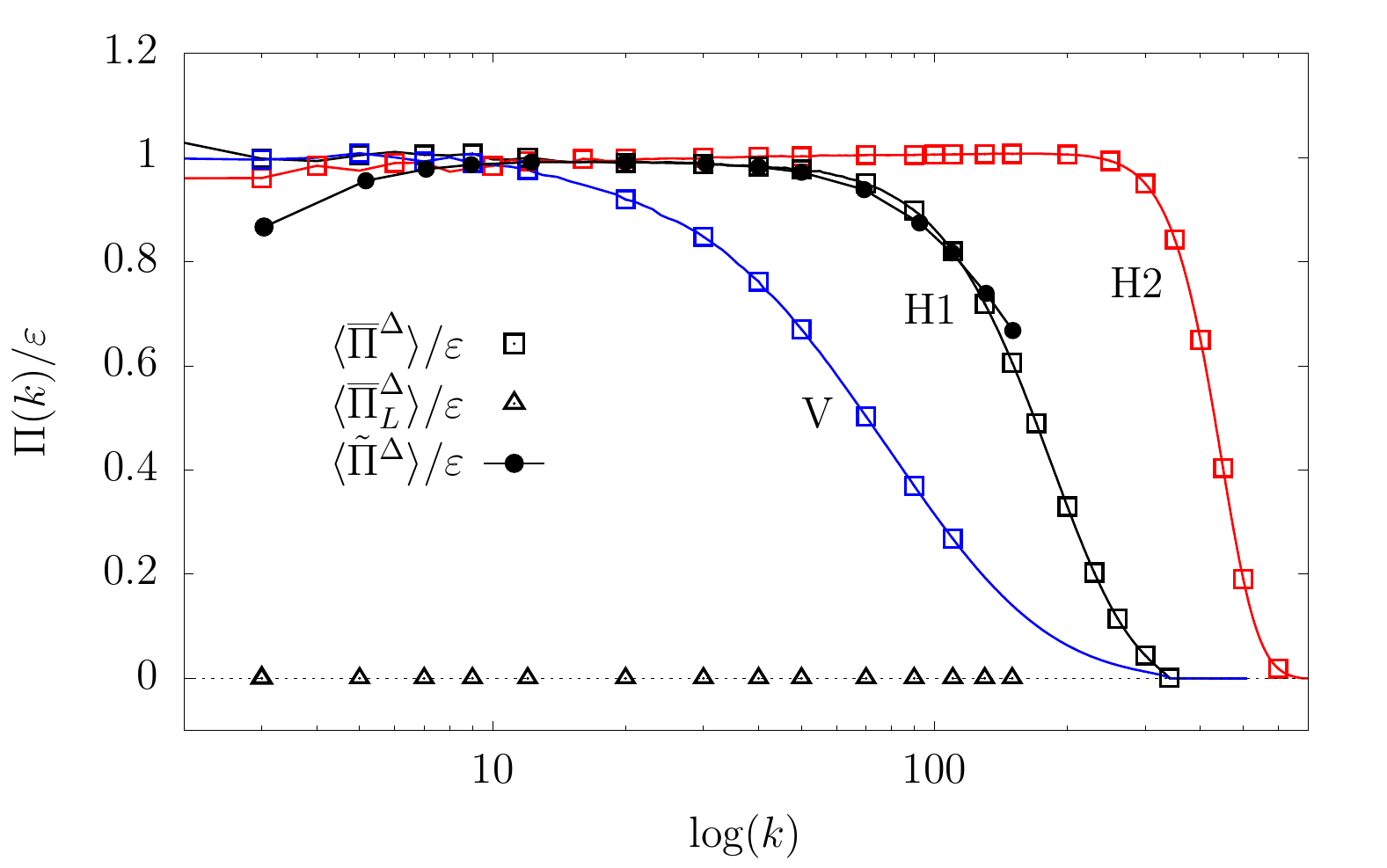}
  \caption{
\mb{Comparison between the normalised Fourier-space energy flux and the 
normalised P-SFS (squares) and F-SFS(dots) energy transfers and the Leonard component (triangles)
as a function of the cut-off wavenumber $k_c = \lb{\pi}/\Delta$ for data-sets V1 (blue/dark grey), V2 (green/light grey),  H1 (black) and H2 (red/grey). The F-SFS energy transfer for the smooth Gaussian filter is shown for data-set H1 only.}}
\label{fig:pi_normalised}
\end{center}
\end{figure}

\noindent
Each data-set consists of a set of \obs{instantaneous velocity fields} sampled  
after the simulations have reached a statistically stationary state.  
The steady-state energy spectrum $E(k)$ and the dissipation spectrum $k^{2}E(k)$ 
\obs{obtained by averaging over the sampled data} for data-sets
V1, V2, H1 and H2 are shown in Figure~\ref{fig:simulations_flux_spectra}(a,b), respectively. 
The effects of using hyperviscosity are different if judged on the spectral properties or on the energy flux. Hyperviscous data have a large bottleneck \cite{Falkovich94} in the high wavenumber range which makes unclear if there is a gain in the scaling range with respect to the viscous case for the same resolution (see Figure~\ref{fig:simulations_flux_spectra}). On the other hand, the extension of the scaling for the hyperviscous energy  flux is more pronounced (see Figure~\ref{fig:pi_normalised}).


\section{Background considerations} \label{sec:theory}  
\obs{As said, in this paper we will focus only on
  {\em a priori} analysis of DNSs of isotropic turbulence.
  The main interest is to have a systematic benchmark of key turbulent statistical properties to validate real applications of LES. In the following, we briefly summarise the main subtleties connected to 
the definition of  filter  in LES and on its impact on the multiscale statistics of the sub-grid energy transfer. }
The governing equations for LES are derived by 
first applying a filtering operation to the incompressible Navier-Stokes equations
\be
\p_t \tbv  + \nabla \cdot (\widetilde{\bv \otimes \bv}) = -\nabla \tP + \nu \Delta \tbv \, 
\label{eq:filtered_nse}
\ee
where the filtered quantities are defined through a filter $G_\Delta$, and $\Delta$ indicates
the filter threshold.  
The filtered velocity field is then given by  
\be
\tbv(\bx,t) \equiv \int_\Omega  d\by \ G_\Delta(|\bx-\by|)\ \bv(\by,t) 
= \sum_{\bk \in \mathbb{Z}^3}  \hat G_\Delta(|\bk|)\ \hat{\bv}(\bk,t) e^{i \bk \bx} \ ,
\ee
with $\hat G_\Delta$ being the Fourier transform of $G_\Delta$.
The aim of LES is to describe the dynamics of the larger scales of the flow, hence the
filtering operation given by $G_\Delta$ is a `coarse-graining' procedure which removes 
scales smaller than the given threshold $\Delta$. In this paper, we will 
use a Gaussian kernel 
$\hat G_\Delta(|\bk|)= \exp(-|\bk|^2\Delta^2/2)$. In order to explicitly separate the terms depending on the SFSs, it is useful to introduce the filtered SFS stress tensor (F-SFS). Note that the filtering operation does not produce a clear Fourier spectral distinction between resolved and unresolved scales; hence in this context, the adjective `resolved' is meant to characterise the energy left to the field at each scale after the filtering operation.
\be
 \ttau_{ij}(\bv,\bv) \equiv  \widetilde{v_iv_j} - \tv_i\tv_j \ , \qquad \text{\obs{(F-SFS)}} \ 
\label{eq:tau_les}
\ee
and to rewrite \obs{Equation~\eqref{eq:filtered_nse}} as 
\be
\p_t \tbv  + \nabla \cdot(\tbv \otimes \tbv) = -\nabla \tP -\nabla \cdot \ttau(\bv,\bv)+ \nu \Delta \tbv \ . 
\label{eq:les}
\ee
\obs{Note} 
that the definition of $\ttau_{ij}$ comes from the exact application of a low
pass filter to each term of the Navier-Stokes equations. So, as \obs{long} 
as 
the unclosed SFS-tensor \obs{is known}, 
the filtering procedure is still exact and  
Equation (\ref{eq:les}) describes the evolution of a filtered field at 
all time. 
In \obs{application,  the filtering protocol would be useless without the introduction of a  closure model for $\ttau(\bv,\bv)$}
\obs{in terms of the resolved-scale velocity,}
i.e.~$\ttau(\bv,\bv) \rightarrow \ttau_{mod}(\tbv,\tbv)$,
\obs{such that}
\be
\p_t \tbv  + \nabla \cdot (\tbv \otimes \tbv) = -\nabla \tP -\nabla \cdot \ttau_{mod}(\tbv,\tbv) \ . \qquad \text{\obs{(F-LES)}} \ 
\label{eq:les-mod}
\ee
\obs{In the following, we will refer to formulation (\ref{eq:les-mod}) as the ``Filtered LES" (F-LES).} 
\obs{The solution to Equation~(\ref{eq:les-mod})} \obs{leads} 
to a break of the property of being a `filtered' field, because the product
of two or more filtered quantities  is not the result of the application of a
filter and both the advection terms on the left-hand side (LHS) and the 
closure for the SFS-tensor on the right-hand side (RHS)
will introduce \obs{uncontrolled} 
errors in any defiltering procedure.
More precisely, since the Gaussian filter retains all scales, it is in principle invertible and 
one could recover the full field from the filtered data. However, replacing the SFS stress tensor 
with a model breaks this property of the filter and leads to errors in the reconstructed data 
\cite{Pope00} (note that defiltering can be an ill-conditioned operation). 
 As a result, any practical implementation of (\ref{eq:les-mod}) will
need to evolve the equations on a numerical grid and make sure to introduce
some effective numerical dissipation scheme that will remove the energy
transferred subgrid. 
Moreover, the original exact filtered
Equation (\ref{eq:les}) does depend on the shape of the filter, while in
(\ref{eq:les-mod}) we have lost connection with the original filtering
protocol, opening the question about how to validate the closure. In principle, one
should reverse-engineer the procedure: given the results of a LES
evolution, derive the filter shape that would give the \obs{correct} agreement if applied to a
 fully resolved evolution \cite{Pope00}. Still, there are problems to define the set of
observables that \obs{should} 
be used to follow this procedure. \\ 

\noindent
A natural way to avoid the above complications is to use a filter which is a
projector, i.e. \obs{a filter that produces the same result when operating 
multiple times on the same field.} 
In terms of its Fourier expression, \obs{a projector filter has the property that:} 
$(\hat G_\Delta(|\bk|))^2 = \hat G_\Delta(|\bk|)$ \cite{Pope00}.  In what follows, the
distinction between projector and non-projector \obs{filters} is reflected in the notation
$\overline{(\cdot)}$ for projected quantities, while $\tilde{(\cdot)}$ is used
to indicate the application of a filter which is not a  projector.
The most common projector widely used in LES for  both real and Fourier applications 
\cite{Pope00,kraichnan1976,MeneveauARFM,chollet1982,Lesieur08}
is a Galerkin truncation for all wavenumbers larger than a given cut-off 
wavenumber 
$k_c \lb{ =\pi}/\Delta$
\be
\obv(\bx,t) \equiv \sum_{\bk \in \mathbb{Z}^3} \hat G_\Delta(|\bk|) \ \hat{\bv}(\bk,t) e^{i \bk \bx} 
= \sum_{|\bk| < k_c } \hat{\bv}(\bk,t) e^{i \bk \bx} \ .
\ee
In order to define 
\obs{the evolution of $\obv(\bx,t)$ properly, i.e. such that it is confined 
to the same finite-dimensional vector space,}
we need to \obs{project} 
the non-linear term of Equation~(\ref{eq:les}), resulting in 
\be
\p_t \obv  + \nabla \cdot(\overline{\obv \otimes \obv}) 
= -\nabla \oP -\nabla \cdot \otau(\bv,\bv)+ \nu \Delta \obv \ . 
\label{eq:Ples}
\ee
In \obs{Equation} 
(\ref{eq:Ples}), the projected SFS \obs{stress} tensor  (P-SFS) is now given by
\be
\label{eq:tau_ples}
 \otau_{ij}(\bv,\bv) =  \overline{v_iv_j} - \overline{\ov_i\ov_j} \ ,  \qquad \text{\obs{(P-SFS)}} \ . 
\ee
\obs{Again, in real \emph{a posteriori} \cite{Piomelli88} implementations, the P-SFS stress tensor in Equation \eqref{eq:tau_ples} 
should be  replaced by a model, i.e. $\otau(\bv,\bv) \rightarrow \otau_{mod}(\obv,\obv)$, 
and one obtains} 
\be
\p_t \obv  + \nabla \cdot (\overline{\obv \otimes \obv}) = -\nabla \oP -\nabla \cdot \otau_{mod}(\obv,\obv) \ , \qquad \text{\obs{(P-LES)}} \ 
\label{eq:Ples-mod}
\ee 
\obs{which in the following is referred to as `Projected LES' (P-LES).} 
Notice that we now have a consistent definition of the `filtering’ protocol and we should have called the unclosed tensor in \eqref{eq:tau_ples} `subgrid-scales’ (SGS) differentiating it from the one called `subfilter-scales’ (SFS) in (\ref{eq:les}) as it can be found in previous literature \cite{Zhou01, carati2001, winckelmans2001}. The difference originates from the fact that the Galerkin truncation removes all Fourier modes below the cut-off scale, while a non-projector filter does not necessarily do so. Hence, unlike for a non-projector filter, for a Galerkin projector there is an exact correspondence between the finest LES grid scale and the cut-off scale. Keeping in mind the above difference, in our {\em a-priori} analyses, the velocity fields are always evolved by fully resolved DNS, inconsistent with the concept of a-posteriori LES on a specified grid. For this reason, we have simplified the notation in F-SFS and P-SFS which allows us to give emphasis to the filters’ properties.

In formulation (\ref{eq:Ples-mod}), 
 the inertial term is also a projected  function, and the evolution of
$\obv(\bx,t)$ is confined to a manifold whose dimension is specified by the
chosen threshold $k_c$ (see the following for other possible definitions of non-sharp Fourier-projectors). 
It may be worth mentioning that, in mathematical analysis, 
one can use a sequence of decreasing filter scales of 
a Galerkin projector to converge to a weak solution of the Navier Stokes equations along a 
subsequence \cite{Doering95}. 
It is important to realise that one might have used a formulation like
 Equations~\eqref{eq:tau_ples} and \eqref{eq:Ples}  for filters
which are not  
projectors. This would not solve the dichotomy among filter-scale and grid-spacing 
and, more importantly,  
double filtering the inertial term with a non-projector filter breaks the 
Galilean invariance of the corresponding SFS stress tensor as shown  
in Appendix \ref{app:GalInv_filter}. 
Using a P-LES has also the advantage (in principle) that the shape of
the filter explicitly appears in the space time evolution, because of the need
to further project the non-linear term (and the modelled \obs{SFS} tensor) at
each time step in (\ref{eq:Ples}). 
\mb{In practice, any numerical implementation of both F-LES or P-LES, requires a dynamical projection on a finite grid which takes into account also the possible aliasing errors. Such a further projection needs always to be applied in the implementation of the F-LES system. For the P-LES system, the projector in front of the non-linear term includes already a dealiasing operation 
if the filter cut-off wavenumber is smaller than $2/3k_{max}$, where $k_{max}$ is the largest resolved wavenumber.}
In any case, aliasing error does not play a role in the {\em a priori} data analysis carried out here, 
since all data were obtained from fully dealiased DNS. This means that the F-LES (6) should be seen as `mathematical LES' and not practical for a-posteriori LES implementations on finite grids \cite{carati2001}.
\mb{The choice of the smooth filter used in our F-LES ensures a high level of filtering after the cut-off scale, implying that the effect of a further projection on the grid would be negligible. This is possible because we considered only the a-priori analysis. In any practical LES implementations, an additional projection on the LES grid and an SGS model with dissipation to maintain stability are needed, even in case of strong filtering.}
\mb{The main mathematical advantage of considering smooth filters is that the SFS stress tensor is positive definite.} As a consequence, they can be applied in the derivation of scaling properties of the physical space velocity field \cite{Eyink05}.

\subsection{\emph{A priori} definition of the energy transfer using the F-SFS or   P-SFS formulations} \label{sec:galilean}
A key benchmark quantity to validate the accuracy of a  LES is the ability to reproduce the correct mean and fluctuating properties of the SFS energy transfer. 
In the smooth filtering approach, where no projectors are applied to the equations, the resolved kinetic energy evolves according to
\be
\frac{1}{2}\p_t(\tv_i \tv_i)   +  \p_j B_j   =  -\tPi 
\label{eq:sg-ene}
\ee
where $B_{j} = \tv_j \frac{\tv_i \tv_i}{2} + \tv_i (\tP \delta_{ij}+
\ttau_{ij})$ is a spatial transport term
that redistributes the resolved energy among
different spatial positions  
while 
\be
\tPi = - \ttau_{ij}(\p_j \tv_i)) = - \ttau_{ij}\tS_{ij} \ , 
\ee
is the instantaneous SFS energy flux, where $\tS_{ij}=(\p_i \tv_j + \p_j \tv_i)/2$ denotes 
the resolved strain-rate tensor.
In the formulation based on a projector filter, the local 
resolved \obs{kinetic} energy evolves
\obs{differently}, since
from Equation~\eqref{eq:Ples} we obtain the following evolution equation for the resolved 
kinetic energy:
\be 
\frac{1}{2}\p_t(\ov_i \ov_i)   =
-\ov_i \overline{\ov_j \p_j \ov_i} - \ov_i \p_j \overline{p}\delta_{ij} - \ov_i \partial_j \otau_{ij} \ ,
\label{eq:Psg-ene} 
\ee 
where the first term on the RHS is
no longer a total derivative, since  
\be \ov_i \overline{\ov_j \p_j \ov_i} = \partial_j
(\ov_i \overline{\ov_j \ov_i} )  - (\p_j \ov_i)( \overline{\ov_j \ov_i} )  \ .
\ee
\obs{The} equation for the SFS energy transfer \obs{therefore} becomes 
\be 
\frac{1}{2}\p_t(\ov_i\ov_i)   +  \p_j A_j   =  -\oPi +  (\p_j \ov_i)(\overline{\ov_j \ov_i} )
\label{eq:PLES-ene} 
\ee 
with $A_{j} = \ov_i ( \overline{\ov_i \ov_j} + \overline{p} \delta_{ij}+ \otau_{ij})$ 
\obs{as the} 
flux term. 
Now, the RHS of Equation ~\eqref{eq:PLES-ene} consists of the P-SFS tensor 
\be
\label{eq:PLES-flux} 
\oPi = - \otau_{ij}\oS_{ij} \ ,  
\ee
and \obs{the}   
additional term 
\be
\opi= (\p_j \ov_i)( \overline{\ov_j \ov_i}) 
\ee 
which  is not Galilean invariant. \obs{Note}  
that in Equations 
(\ref{eq:sg-ene}), (\ref{eq:Psg-ene}) and \eqref{eq:PLES-ene} 
we have not explicitly written, for the sake of simplicity, 
the viscous contributions. 
We will omit the \lb{viscous}  
terms in the remainder of this paper.  
\\

\begin{figure}[H]
\begin{center}
 \advance\leftskip-1.6cm
  \includegraphics[scale = 0.55]{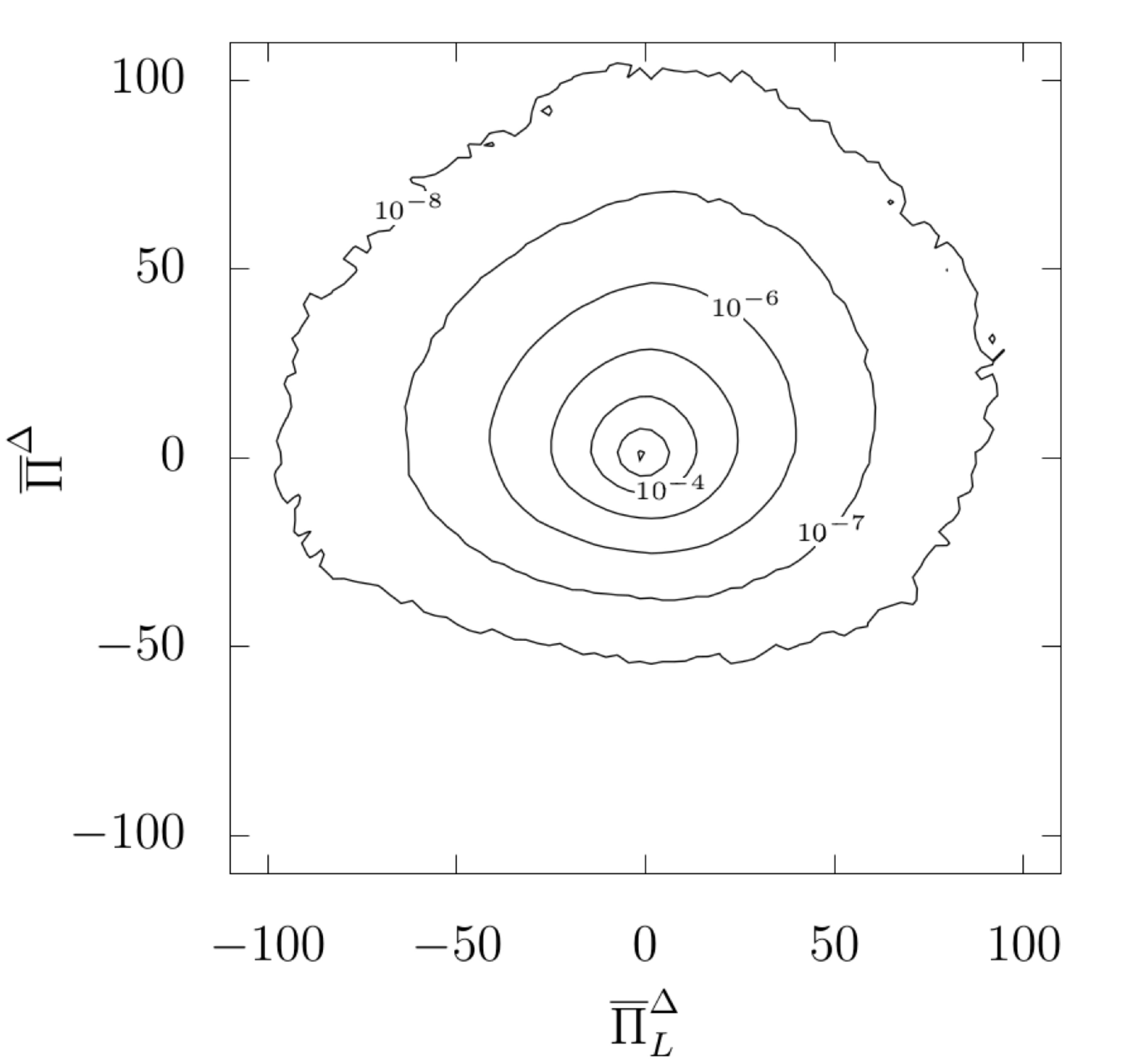}
  \caption{Joint pdf  
           of the P-SFS energy transfer and the Leonard component 
           for a sharp projector at
           cut-off wavenumber $k_c=20$ from data-set H1. 
}
\label{fig:jointpdf_leon_plesgal}
\end{center}
\end{figure}

\noindent
The lack of Galilean invariance of $\opi$ does not break 
the {\em global} Galilean invariance of Equation~\eqref{eq:PLES-ene} because of cancellations with terms on the LHS (see 
Appendix~\ref{app:GalInv}). Nevertheless,  it is clear that adopting this formulation,  the total  
\obs{SFS} energy transfer  using the P-SFS stress tensor:
\begin{equation}
\label{eq:totalP}
\mathcal{P}^\Delta=-\oPi+\opi = -(\p_i\ov_j)\overline{v_iv_j}
\end{equation}
 is not pointwise  Galilean invariant. The non-Galilean invariant term,  $\opi$,
is closed in terms of the resolved fields and must not be considered 
a true SFS transfer. Indeed, it is easy to realise that its
mean value over the whole volume is always vanishing: 
\begin{align}
\langle \opi \rangle =
\frac{1}{|\Omega|}\int_\Omega  d\bx \ (\p_j \ov_i) \overline{\ov_j \ov_i} = 0 \ , 
\end{align}
where we have  used the filter property
$ \left \langle f \overline{g} \right \rangle = \left \langle \overline{f} g\right \rangle$,
the projector property 
$G^2=G$
and incompressibility to write $(\p_j \ov_i) \overline{\ov_j \ov_i}$ as a total derivative.
The \obs{net}  
SFS energy transfer $\mathcal{P}^\Delta$ consists of genuine SGS 
coupling term 
$\oPi$ and a contribution \obs{$\opi$} due to self-coupling of the resolved scales
which 
breaks pointwise Galilean invariance.
The  pointwise lack of Galilean invariance of the `unsubtracted flux'  
 results in unphysical 
large fluctuations as shown in  \cite{aluie2009II} which  might  lead to different multiscale results   compared to those of $\oPi$ and $\tPi$ (see Section~\ref{sec:num_galilean}). 
\\
\noindent
The lack of Galilean invariance  can be solved by exploiting the 
freedom to add and subtract a term 
that will make both the RHS and the LHS separately  Galilean invariant. 
In particular,  we  rewrite the energy balance as
\begin{align}
\frac{1}{2}\p_t(\ov_i \ov_i)   +  \p_j A_j  =  -\overline{\Pi} +  (\p_j \ov_i)\tauleo + \frac{1}{2}\p_j(\ov_j \ov_i\ov_i)  \ ,   
\end{align}
where we have introduced the Leonard stress \cite{Leonard75}, $\tauleo$:  
\be
\label{eq:corrector}
\tauleo \equiv \overline{\ov_i\ov_j}-\ov_i\ov_j \ , 
\ee
plus another term which is a total derivative and that can  be moved to the LHS of Equation~\eqref{eq:PLES-ene}.
Hence, the kinetic energy balance based on the P-SFS stress tensor becomes
\be
\label{eq:sg-eneP-leo}
\frac{1}{2}\p_t(\ov_i \ov_i) + \p_j \left(A_{ij} - \frac{1}{2}\ov_i\ov_i \ov_j\right)   =  -\oPi -\Pileo \ ,
\ee
where $\Pileo = -\tauleo\oS_{ij}$ is the energy transfer corresponding to the Leonard stress.
The RHS and LHS of Equation~\eqref{eq:sg-eneP-leo} are now separately 
Galilean invariant, 
 and we have a way to assess the properties of the energy balance without being affected
by spurious effects.
The introduction of the Leonard stress tensor has been also used in the literature to preserve Galilean invariance in the definition of models for a-posteriori LES \cite{thiry2016}.
 \begin{table}
 \begin{center}
 \begin{tabular}{clll}
  ID & Stress \lb{tensors} &  & Energy transfer\lb{s}   \\
  \hline
 F-SFS & $\ttau_{ij}=\widetilde{v_iv_j} - \tv_i\tv_j$ & \lb{} & $\tilde{\Pi} = -\obs{\tS_{ij}}\ttau_{ij}$ \\
 P-SFS & $\otau_{ij} =  \overline{v_iv_j} - \overline{\ov_i\ov_j}$ & \lb{} & $\overline{\Pi} = -\obs{\oS_{ij}}\otau_{ij}$ \\
 P-SFS(nG) & $\otau_{ij} =  \overline{v_iv_j} - \overline{\ov_i\ov_j}$ &  \lb{} & $\mathcal{P}^\Delta = -\obs{\oS_{ij}}\overline{v_iv_j}$ \\ 
 Leonard & $\tauleo =  \overline{\ov_i \ov_j}-\ov_i\ov_j$ &  \lb{} & $\Pileo = -\obs{\oS_{ij}}\tauleo$ \\
  \hline
  \end{tabular}
  \end{center}
 \caption{Summary of definitions of stresses and instantaneous energy transfers. The
          label nG indicates that the corresponding SFS energy transfer is not Galilean 
          invariant. 
 }
 \label{tbl:definitions}
 \end{table}
\noindent
The transfer \obs{involving}
SFS quantities is given by $\oPi$ alone because
the Leonard stress describes only a coupling among resolved scales 
and its  the contribution to the mean SFS energy transfer vanishes when averaged on the whole volume.
In Figure~\ref{fig:jointpdf_leon_plesgal}, we show the joint probability density function (pdf) of the two contributions to the total SFS energy transfer  for a typical cut-off in the inertial range, $k_c = 20$,
from where it is clear that  $\oPi$ and $\Pileo$ are \obs{nearly}  
uncorrelated. \\
In summary, the use of a projector to define the filtering operation  results in a sharp distinction between two SFS
energy transfer contributions, one described by a genuine 
correlation among resolved and SFSs, $\oPi$, and another one due to a  self-coupling of
the resolved scales, $\Pileo$. 
On the other hand,  the total 
F-SFS energy transfer $\tPi$ contains local contributions which are not strictly 
associated with the SFSs, since it has the same formal structure as  $\oPi + \Pileo$.
Only the  global average on the whole volume of the  SFS energy transfer is
correctly described by the F-SFS approach,  while its local  values $\tPi$ are
affected  by contributions coming from 
\obs{self-coupling of the resolved scales.}
A summary of all definitions of SFS stresses and energy transfers is given in
Table~\ref{tbl:definitions}.

\section{Anomalous scaling properties of the SFS energy transfer} 
\label{sec:intermittency}
In order to build models which are able to describe 
higher-order statistical features of a turbulent flow, 
it is important to distinguish 
\obs{physically relevant fluctuations from unphysical fluctuations.}
\obs{Unphysical}
fluctuations can be induced by the filter, e.g.  a sharp Galerkin projector 
is discontinuous in Fourier space and therefore induces Gibbs 
oscillations in physical space, which can contaminate 
the measured statistical signal \cite{Ray11}.  
Fluctuations can also originate from the 
residual self-coupling of the resolved scales and from a lack of 
Galilean invariance of the SFS energy transfer, as 
discussed in further detail in the coming section.
Obtaining a clear statistical signature of the SFS energy transfer fluctuations 
is especially important for the assessment of 
backscatter contributions.  
It is known that  the pdf of $\tPi$ obtained 
using a smooth Gaussian filter has large tails skewed toward the positive values, in agreement 
with the existence of a direct \obs{energy} 
cascade both in mean and for local intense events 
\cite{Cerutti98,MeneveauARFM}. 
\subsection{Mean properties}
Figure \ref{fig:pi_normalised} shows a comparison 
between the total flux across a spherical shell of radius $k_c$ in Fourier-space, 
\be
\Pi(k_c) = \sum_{k'=1}^{k_c} \sum_{|\bk|=k'}  ik_j \hat{u}_i(\bk)^*\sum_{\bp \in \mathbb{Z}^3} \ \hat{u}_i(\bp)\hat{u}_j(\bk-\bp) ,
\ee  
and the different contributions  to  the SFS energy transfer for a sharp projector 
$\langle\oPi \rangle$ and $\langle\Pileo \rangle$  
as a function of the cut-off wavenumber $k_c = \lb{\pi}/\Delta$ and for different Reynolds numbers. 
As expected, the contribution originating from 
the Leonard stress vanishes, $\langle\Pileo \rangle = 0$, while the Fourier flux is 
exactly reproduced by $\langle\oPi \rangle$ due to the Parseval identity.
Here and hereafter we adopt the notation $\langle \cdot \rangle$ to indicate an average 
\obs{over the entire} 
physical volume. 
For comparison, we also show the mean SFS energy transfer obtained from the 
Gaussian filter as a function of $k_c = \lb{\pi}/\Delta$. The latter does not match exactly  
 the Fourier-space energy flux at small filter thresholds, indicating non-trivial coupling among degrees of freedoms above and below the filter width.  
Deviations between $\langle\tPi \rangle$ and $\Pi(k)$ can indeed be expected, since the $\langle\tPi \rangle$
can be expressed as weighted average in Fourier-space centred around $k_c = \lb{\pi}/\Delta$ \cite{Eyink05,Rivera14}, 
and there is no {\em a priori} reason for $\Pi(k)$ to match its weighted average. 

\subsection{Effects of non-Galilean invariance} \label{sec:num_galilean}

\begin{figure}[H]
\begin{center}
 \advance\leftskip-0.0cm
  \includegraphics[scale = 0.44]{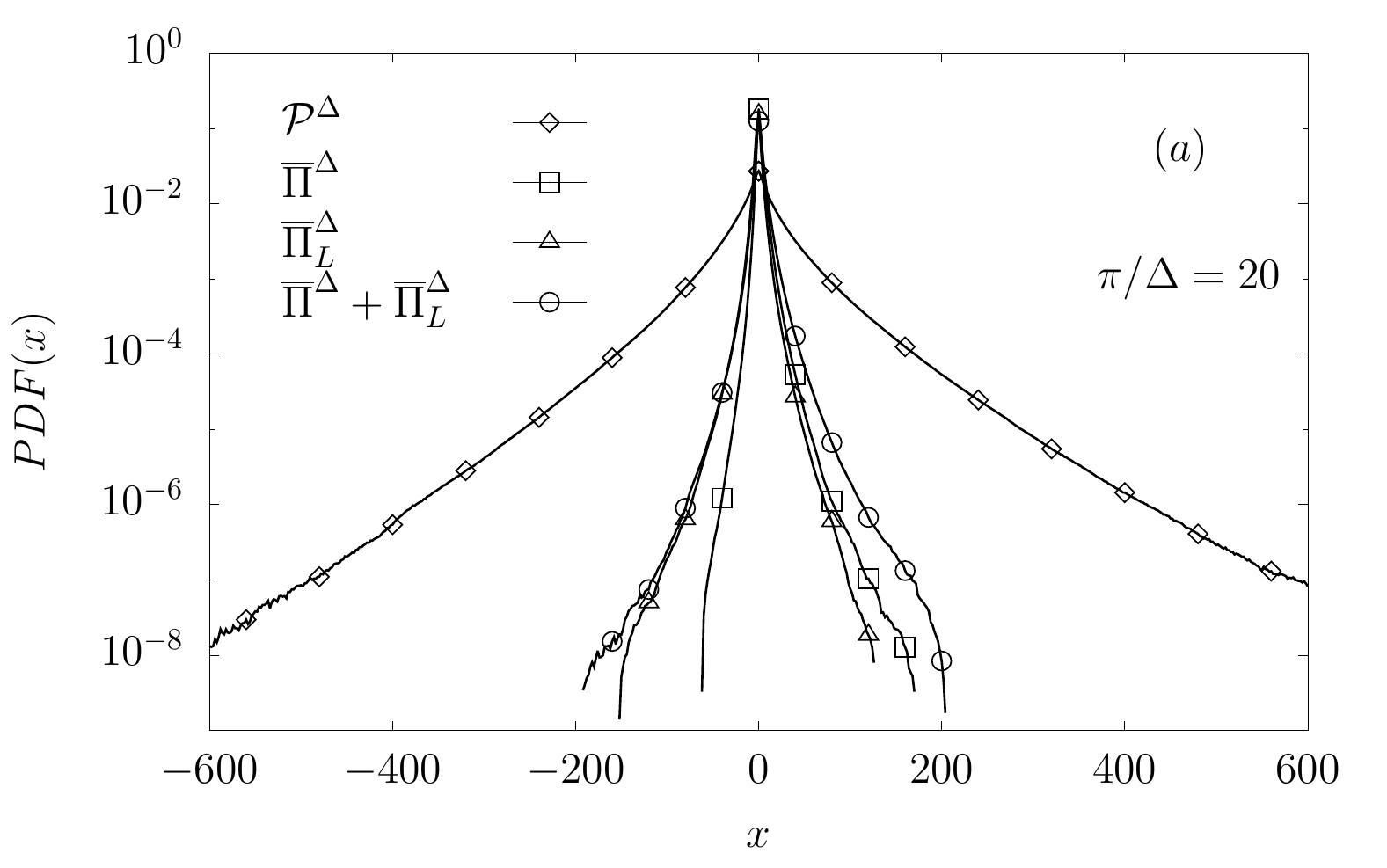}
  \includegraphics[scale = 0.44]{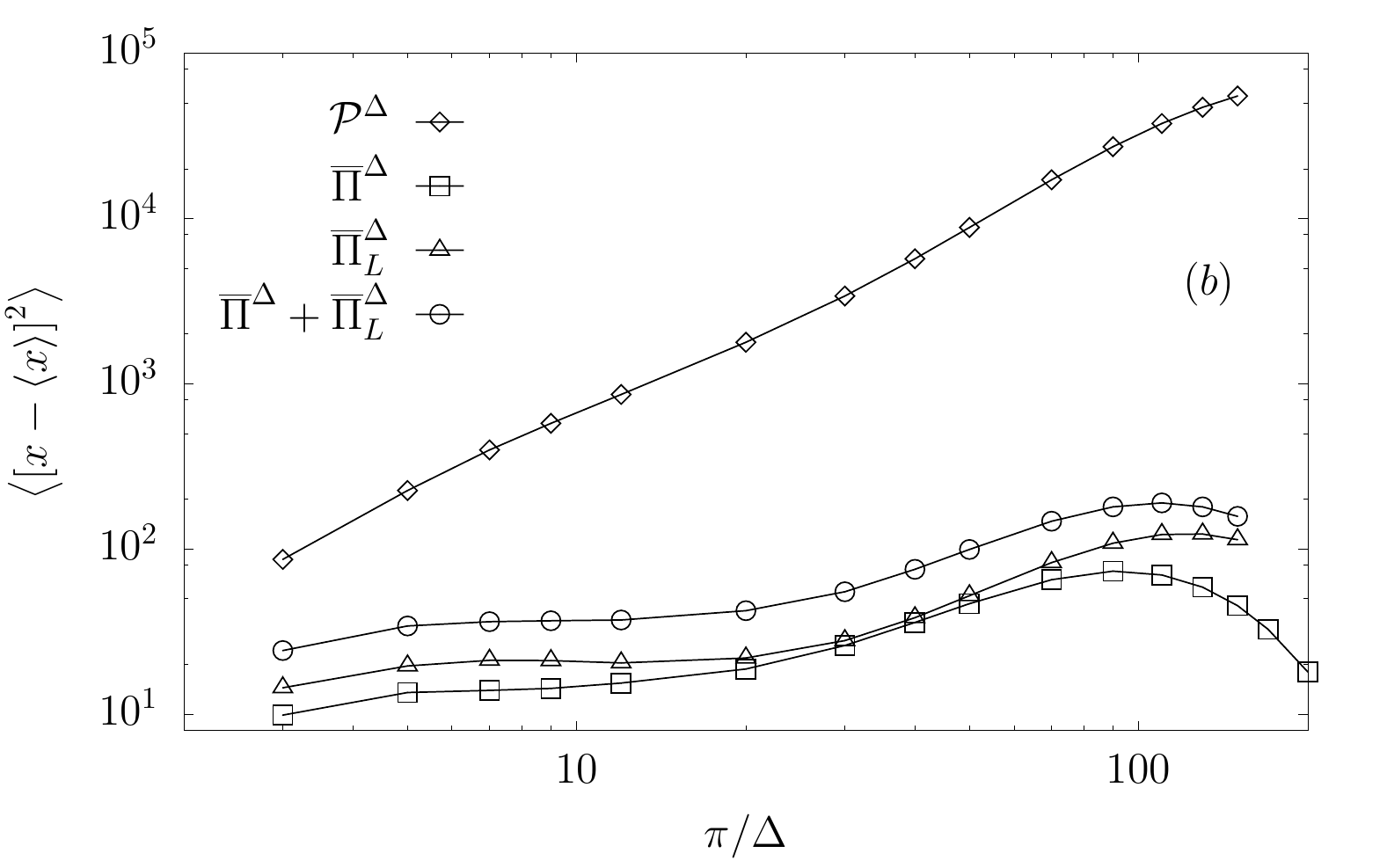} \\ 
  \includegraphics[scale = 0.7]{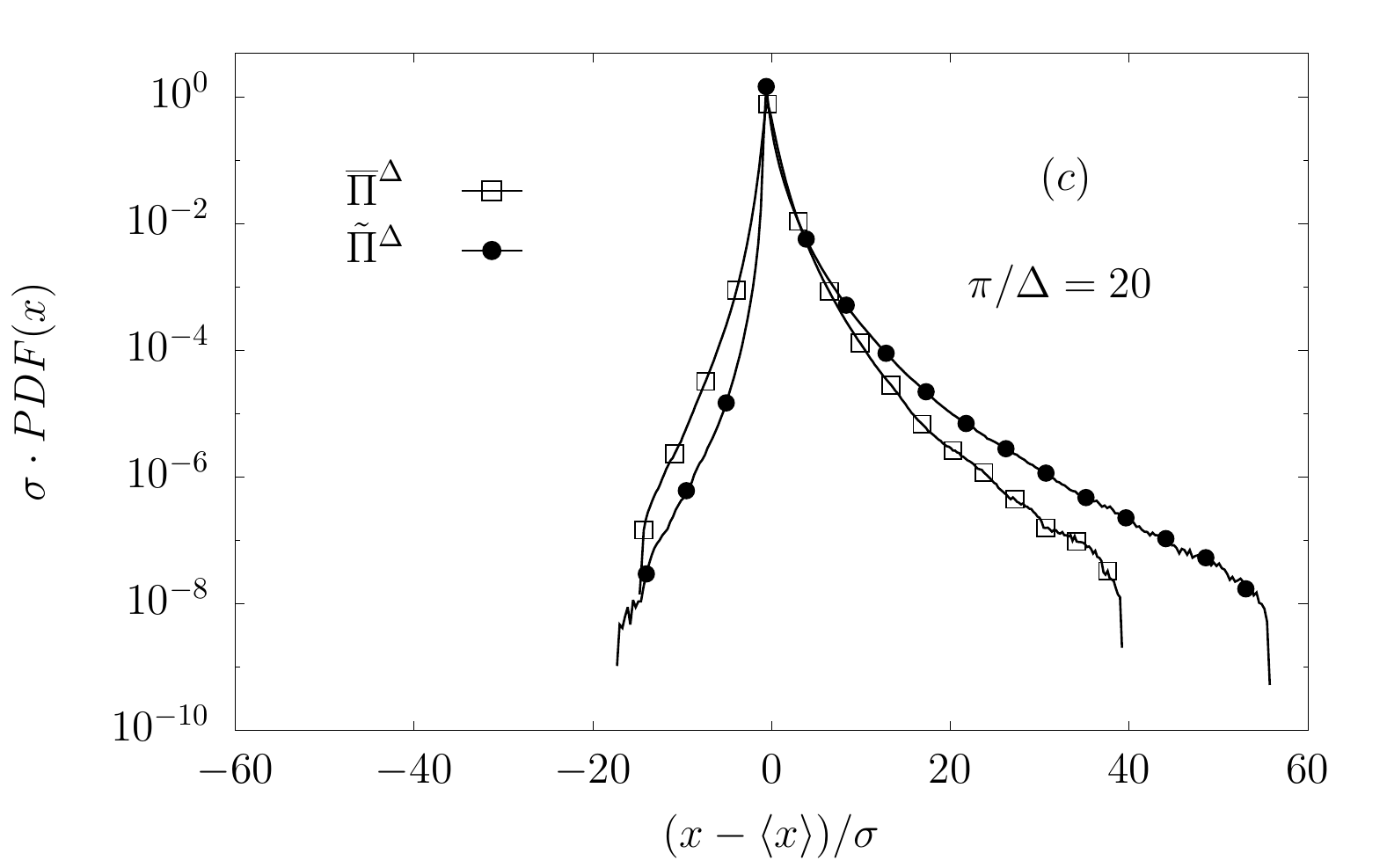}
  \caption{
          (a) The pdfs 
              of different components of the P-SFS energy transfer.
          (b) Standard deviation of the corresponding pdfs
             as a function of the cut-off wavenumber $\lb{\pi}/\Delta$.
         (c) Comparison between the standardised pdfs of the P-SFS energy transfer for
             a sharp projector and of the F-SFS energy transfer for
             a smooth Gaussian filter at $\Delta = \lb{\pi/20}$. Data are taken from data-set H1.
}
\label{fig:pdfs_sigmasq}
\end{center}
\end{figure}

According to Equation~\eqref{eq:PLES-ene}, the non-Galilean invariant definition of the
P-SFS energy transfer $\mathcal{P}^\Delta$ consists of two terms: $\oPi$, which is
Galilean invariant and couples the resolved and the unresolved scales 
and the non-Galilean invariant $\opi$, which is given only in terms of the resolved fields.
In  Figure~\ref{fig:pdfs_sigmasq}, we show the effects of breaking Galilean invariance by comparing the pdf of the
different contributions and of the ones obtained by introducing the Leonard stress and therefore recovering the invariance term by term. As one can see, the fluctuations of the `unsubtracted flux' $\mathcal{P}^\Delta$ 
are much larger than those of the invariant terms, confirming the importance of Galilean invariance
\cite{aluie2009I,aluie2009II}.
In order to quantify the difference between the fluctuations 
of $\mathcal{P}^\Delta$ and the other components of the P-SFS energy transfer, 
we also show the standard deviations of all components.
In Figure~\ref{fig:pdfs_sigmasq}(b), we show that the  non-Galilean invariant 
definition of the P-SFS energy transfer $\mathcal{P}^\Delta$ is between one and three
orders of magnitude larger than those corresponding to the other terms. 
\\
Since the fluctuations of $\oPi$ are orders of magnitude smaller than the fluctuations
of $\mathcal{P}^\Delta = \oPi + \opi$, the large tails of the latter  must be connected to $\opi$.  The
question now arises whether these large spurious fluctuations originate from a
lack of Galilean invariance or if they are due to 
the self-coupling among the resolved scales.  The latter can be quantified through the fluctuations of the 
component of the Leonard stress,  $\Pileo$. As shown by 
Figure~\ref{fig:pdfs_sigmasq}(a,b), both  $\oPi$ and $\Pileo$
have similar fluctuations.  We therefore conclude that a lack of Galilean invariance has a drastic effect
on the fluctuations of the SFS energy transfer and that it becomes larger and larger by decreasing the cut-off scale.  
\\
\noindent
The properties of the non-Galilean invariant `unsubtracted flux' $\mathcal{P}^\Delta$ 
obtained from a sharp spectral projector were discussed in Ref.~\cite{aluie2009II}
also in the context of locality of the energy cascade. Unlike $\oPi + \Pileo$, which was rigorously 
proven to be pointwise scale-local as non-local contributions from sweeping effects were removed, 
$\mathcal{P}^\Delta$ only becomes scale-local in an average sense since the sweeping 
contributions cancel under space averaging. 
\obs{We conclude that potential spurious fluctuations introduced by a lack of Galilean invariance may be of concern in LES,}
and the unsubtracted flux $\mathcal{P}^\Delta$
will not be considered further in this paper. \\

\begin{figure}[H]
\begin{center}
 \advance\leftskip-0.1cm
  \includegraphics[scale = 0.38]{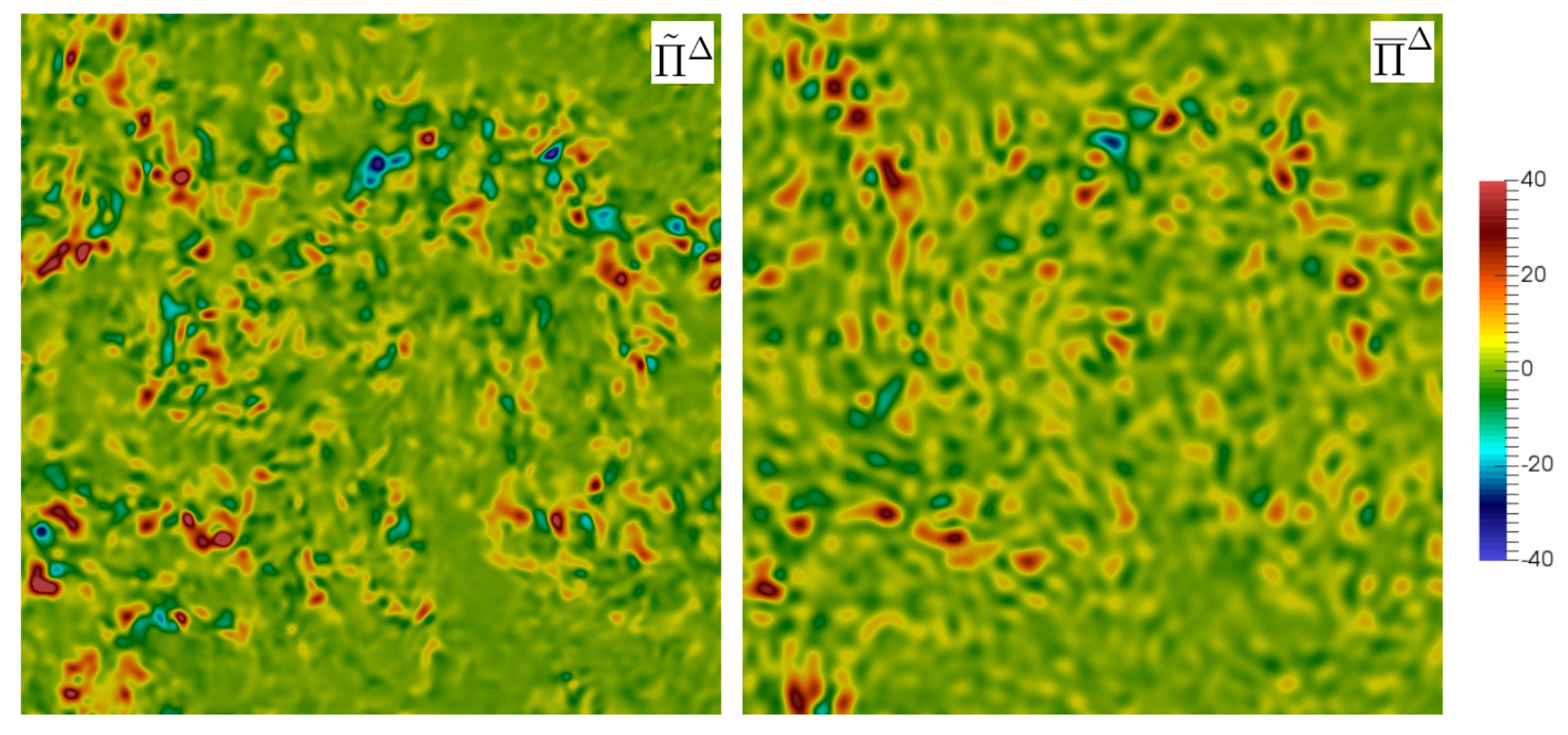}
  \caption{Visualisations of the SFS energy transfer in a plane cut though the volume.
   Positive values indicate forward energy transfer, while negative values correspond to backscatter.
Left: F-SFS (Gaussian filter), right: P-SFS (sharp spectral projector) at $k_c = \lb{\pi}/\Delta = 20$ for data-set H1.
}
\label{fig:visualisation}
\end{center}
\end{figure}
\noindent


In  Figure~\ref{fig:pdfs_sigmasq}(c), we show  the 
normalised pdfs of $\tPi$ and $\oPi$ at comparable filter thresholds. 
Hence, both the Gaussian filter and the real subfilter component of the sharp projector have a
 statistical signal correlated with the global forward energy cascade mechanism. 
This is apparently not the case  for the other components of the SFS energy transfer
obtained from a sharp projector, since the pdfs 
of $\oPi+\Pileo$ and $\Pileo$ shown in Figure~\ref{fig:pdfs_sigmasq}(a) 
are  more symmetric than the pdf of $\oPi$.
Since $\Pileo$ describes the energy transfers among the resolved scales which vanish on average, the negative 
tail in $\Pileo$ and therefore to some extent also the negative tail in 
$\oPi+\Pileo$ cannot be a genuine backscatter signal.
In summary, the above results further support 
the role of  $\oPi$ as the most 
relevant component of the P-SFS energy transfer. 
Figure \ref{fig:visualisation} shows  
visualisations of the P-SFS energy transfer 
$\oPi$ and the F-SFS energy transfer $\tPi$ for data-set H1 obtained by 
a plane cut through the volume. The filter thresholds 
for $\oPi$ and $\tPi$ \lb{are} the same as that of their 
pdfs shown in Figure~\ref{fig:pdfs_sigmasq}(c).    
The data corresponding to the F-SFS energy transfer
resolve slightly smaller structures compared to the P-SFS data, which can be expected as
Gaussian smoothing does not result in a reduction of the degrees of freedom in the same 
way Galerkin
truncation does.  

\subsection{Intermittency and anomalous scaling}
\begin{figure}[H]
\begin{center}
  \includegraphics[scale = 0.7]{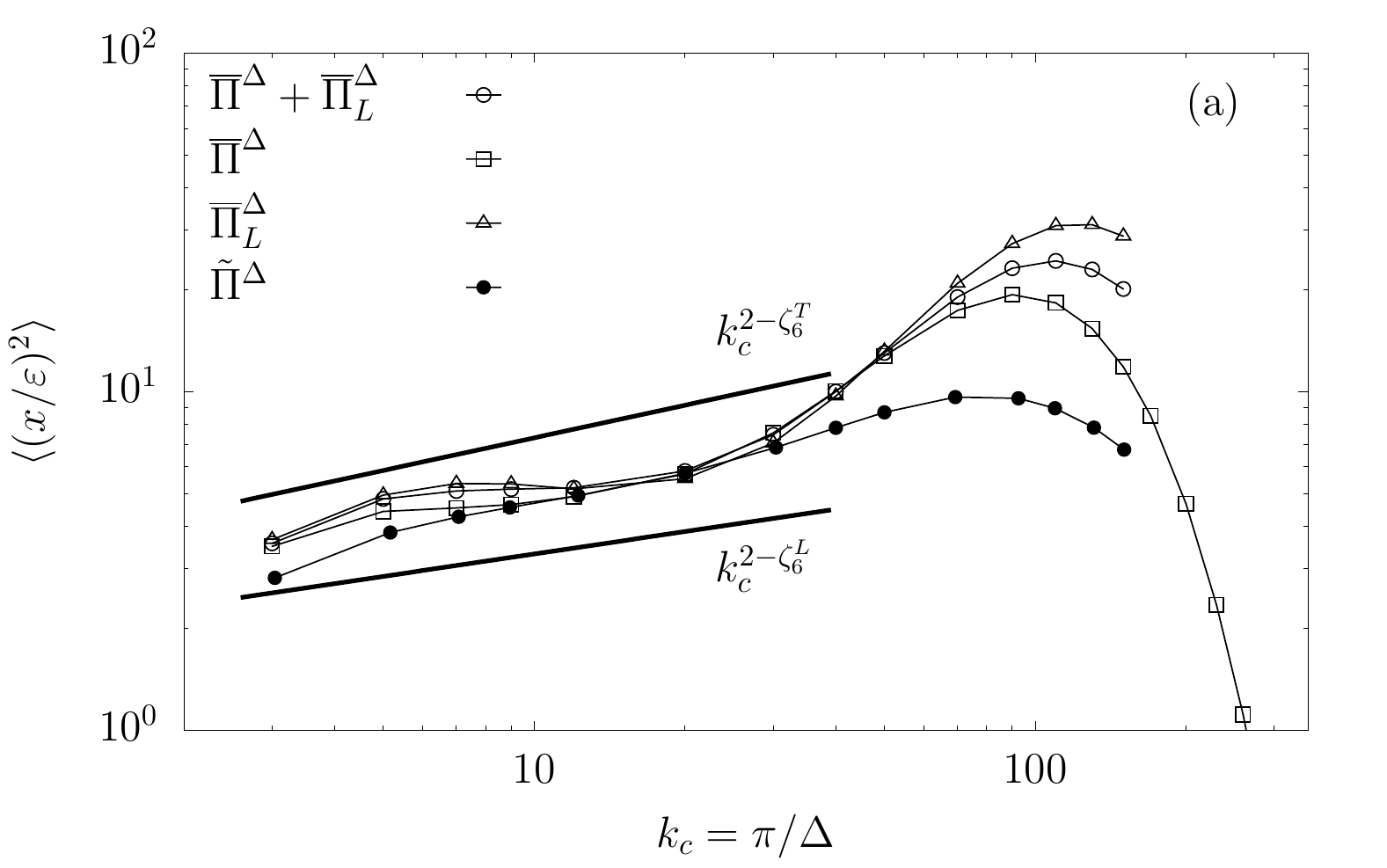} \\
  \includegraphics[scale = 0.7]{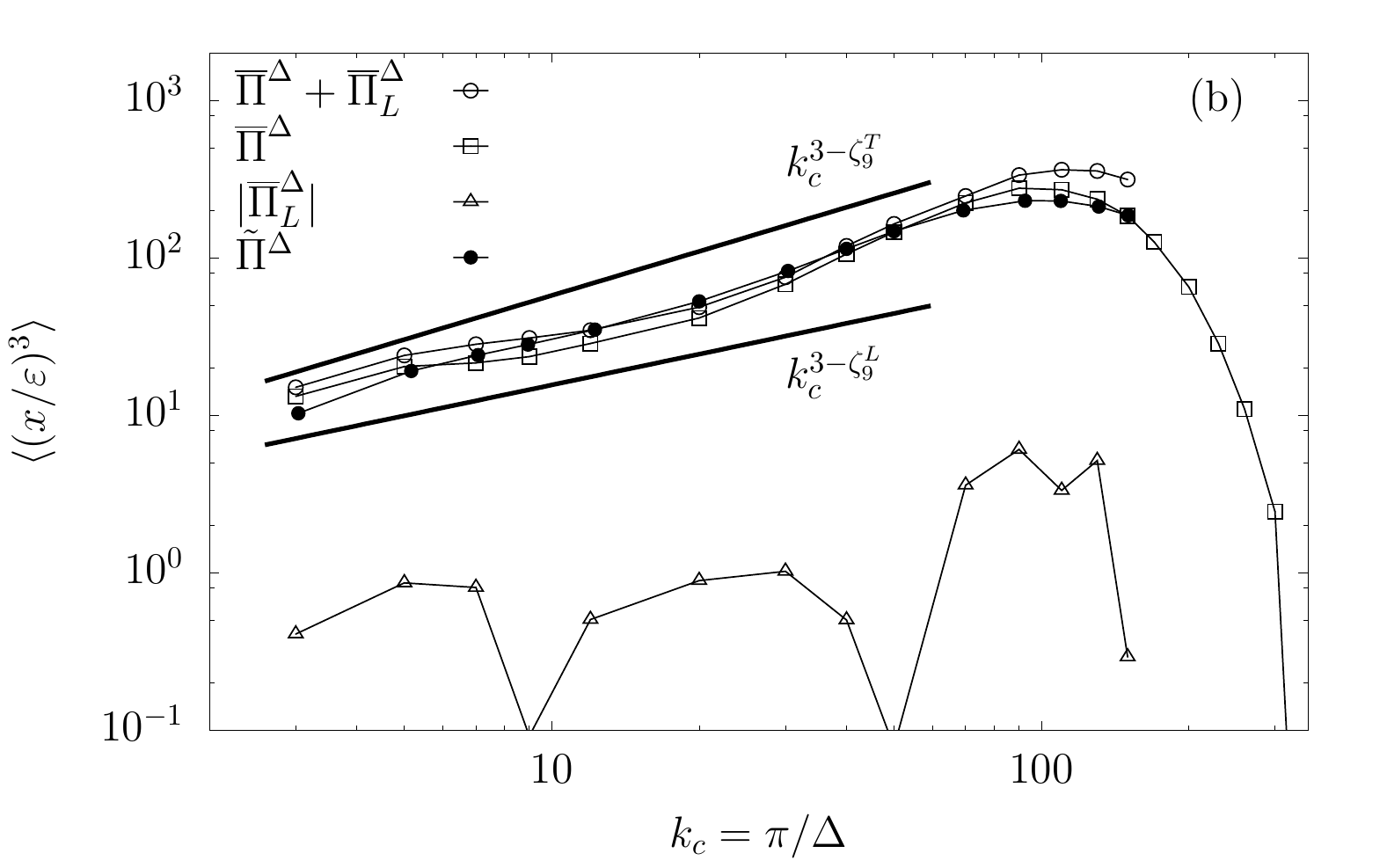}\\ 
  \includegraphics[scale = 0.7]{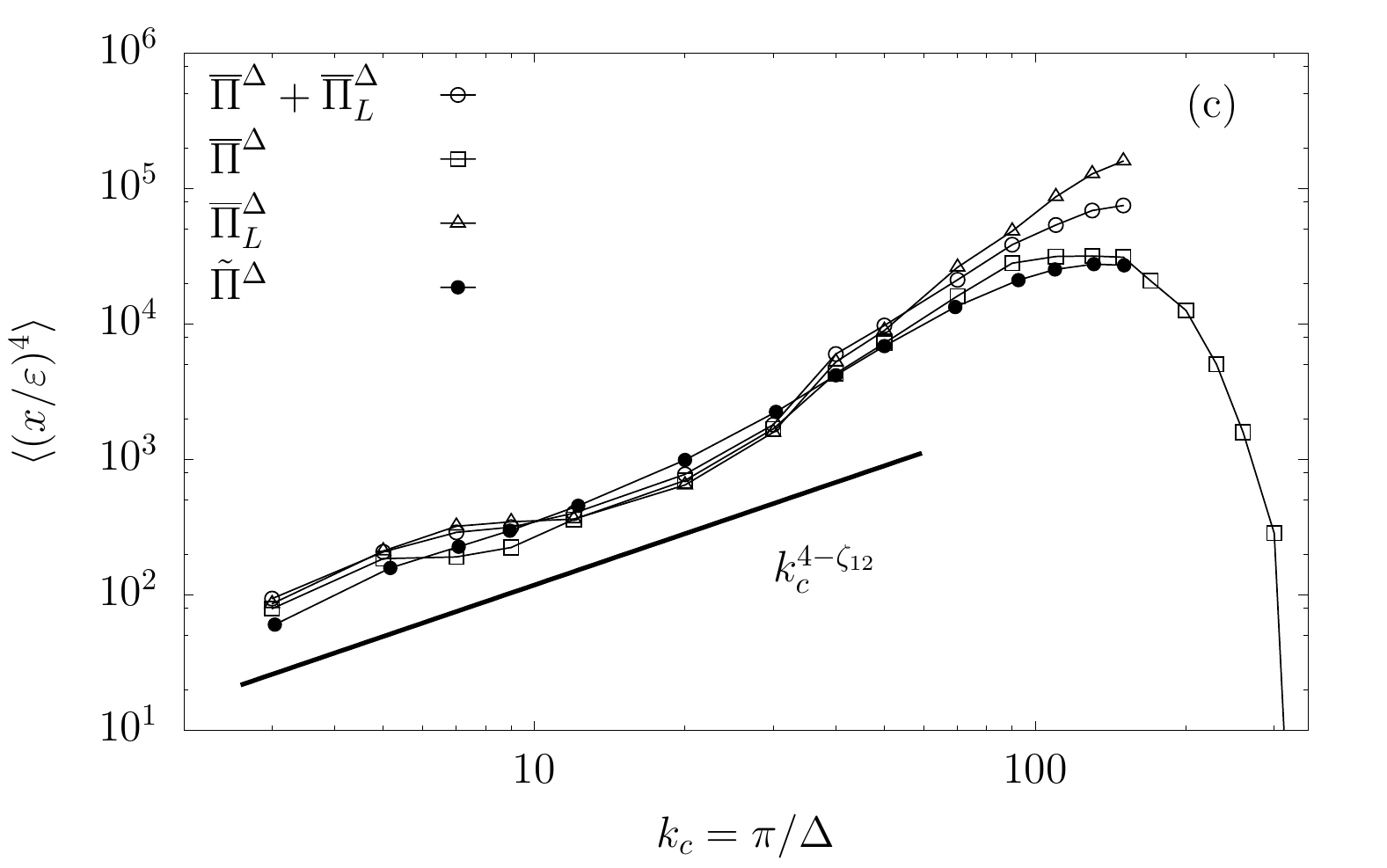}
  \caption{  data-set H1.
             Scaling of the $n^{\textrm{th}}$ moments of
             the different components of P-SFS and F-SFS energy transfers:
         (a) $n=2$,
         (b) $n=3$,
         (c) $n=4$.
          The solid lines indicate the scaling expected from the
          multifractal model and Equation~\eqref{eq:interm_scaling}
          using the anomalous exponents for the
          longitudinal and transverse structure functions $\zeta_{3n}^L$ and $\zeta_{3n}^T$ Ref.~\cite{Gotoh02}. 
          In (c), for $n=4$, the solid line indicates the
          prediction from the She L\'ev\^eque model $\zeta_{3n} = 2.74$ \cite{She94,Boffetta08} is shown.
          The data corresponding to $\oPi+\Pileo$ and $\tPi$ have been
          shifted for presentational reasons.
}
\label{fig:pi_sq_cbH1}
\end{center}
\end{figure}

\obs{An accurate LES should}
reproduce the correct multiscale properties of the SFS stress
tensor and energy transfer. It is well known that turbulence \obs{contains} 
anomalous scaling (intermittency) in the inertial range of scales
\cite{Frisch95}. Intermittency is typically  \obs{measured through}  
the scaling
properties of high-order moments of the velocity increments  as a function of
the separation scale or in terms of moments of velocity gradients  as a
function of Reynolds \obs{number}.  More precisely, the longitudinal and transverse velocity
increments are defined as
$\dru_L (\ux,\ur) \equiv  \druvec(\ux,\ur)\cdot {\urh}$ and 
$ \druvec_T (\ux,\ur) \equiv \druvec(\ux,\ur)-\dru_L (\ux,\ur)\urh $ respectively, where 
$\druvec (\ux,\ur)  \equiv \bv(\ux+\ur)-\bv(\ux)$ is the two-point velocity difference
at separation vector $\ur$ and
$\urh$ is the unit vector along $\ur$. The $n^{\textrm{th}}$-order
moments \obs{of $\dru_L (\ux,\ur)$ and $ \druvec_T (\ux,\ur)$} are the longitudinal structure function (LSF) and
\obs{the} transverse structure function (TSF),
\be
\label{lsf.eq}
S^{(n)}_L(r) \equiv \la(\dru_L (\ux,\ur) )^n \ra \;, \qquad
S^{(n)}_T(r) \equiv \la |\druvec_T (\ux,\ur)|^{n} \ra \;,
\ee
\obs{respectively},
where $\la \cdot \ra$ denotes space and time averages and we have assumed isotropy for simplicity. At high Reynolds \obs{numbers} both
\obs{the} 
LSF and TSF show inertial-range anomalous scaling:
\be
S^{(n)}_L(r) \sim r^{\zeta_L(n)} \qquad S^{(n)}_T(r) \sim r^{\zeta_T(n)} 
\ee
with scaling exponents that are multifractal \cite{Frisch95} and  different from the Kolmogorov prediction $n/3$.  
We differentiate \obs{between} 
longitudinal and transverse exponents because empirical measurements show a small difference \obs{between} 
the two sets (see \cite{Iyer17} for a recent discussion on the Reynolds \obs{number} dependency of the mismatch among 
longitudinal and transverse  scaling exponents and \cite{Biferale05}
 for theoretical considerations). 
A typical signature of intermittency  is given by the growth of the \obs{f}latness: 
\be
F_{L,T}(r) = \frac{S^{(4)}_{L,T}(r)}{(S_{L,T}^{(2)}(r))^2} \sim r^{\zeta_{L,T}(4)-2\zeta_{L,T}(2)}.
\ee
\obs{Because $F_{L,T}=3$ for Gaussian distributions, the}  
empirical observation that $\zeta(4) \neq 2 \zeta(2)$ 
\obs{quantifies}  
the departure 
from Gaussian statistics. 
More importantly, such non-Gaussian fluctuations 
are present even at \obs{relatively} 
small Reynolds numbers \cite{Benzi93,Benzi95,Schumacher14}. 
As a result, the problem of validating any SFS model beyond second-order (spectral) properties \obs{is important,}
 both for applications and fundamental studies. \\
\obs{Of interest is the connection between}
the scaling properties of the SFS energy transfer as a function of the
cut-off $\Delta$ 
\obs{and the scalings}
of the LSF \obs{and} 
TSF as a function of the increment $r$. \obs{One approach is to treat} 
the filter as a local operation in 
\obs{scale space}  
and \obs{to relate} 
the SFS energy transfer at $\Delta$ \obs{to}  
the corresponding dimensional equivalent in terms of 
velocity increments at scale $r=\Delta$:
$ \tPi \sim (\delta_\Delta v)^3/\Delta. $ 
Indeed, at a given filter scale, the SFS-stress tensors $\ttau$ and $\otau$
can be expressed in terms of averages over velocity field increments at scales less than 
$\Delta$ \cite{Constantin94,eyink2,Vreman94a}
\be
\ttau_{ij}(\bx) = \langle \delta_{\br} v_i(\bx)  \delta_{\br} v_j(\bx)\rangle_\Delta - 
                 \langle \delta_{\br} v_i(\bx) \rangle_\Delta  \langle \delta_{\br} v_j(\bx)\rangle_\Delta \ ,
\ee 
where $\langle f \rangle_\Delta =  \int d\br f(\br) G_\Delta(\br)$ denotes a  weighted 
average over the displacement $\br$. The same expression holds for $\otau$.
For H\"older-continuous velocity fields with H\"older-exponent $h$,
i.e. if $|\delta_{\br} \bv(\bx)| = O(|\br|^h)$, the following {\em pointwise} upper bound can be derived 
for the SFS-energy transfer \cite{eyink2}:
\be
\label{eq:local_estimate}
\tPi = O(\Delta^{3h-1}) \ ,
\ee
provided the filter and its gradient are bounded and decrease sufficiently {rapidly} 
at infinity.  
In order to account for the existence of  a multifractal scaling  with different 
local H\"older exponents, the same approach leads \obs{to} a {\em global} upper bound and hence a scaling
estimate     
\be
\label{eq:interm_scaling}
\langle |\tPi|^n \rangle = O(\Delta^{\zeta_{3n}-n}) \ ,
\ee
where $\zeta_n$ are the anomalous exponents of the $n^{\textrm{th}}$-order structure functions \cite{Eyink05} and where we have neglected the
small 
empirically observed \obs{mismatch between} 
longitudinal and transverse increments (which cannot be captured by the above estimate).
As for the local upper bound given by inequality \eqref{eq:local_estimate}, the derivation of the 
rigorous global scaling result presented in Ref.~\cite{Eyink05} requires conditions on the filter functions 
which are not satisfied by generic projector filters.  
However, as we explain in Appendix B, Eyink's \cite{Eyink05} scaling estimates (and upper
bounds) can also be shown to apply to our P-SFS energy transfer in 
Equation~\eqref{eq:PLES-flux}, if we use a smooth filter. Such a smooth filter can be chosen to
approximate a Galerkin projector with arbitrary accuracy at the expense of the
upper bound becoming arbitrarily large. To supplement this, we show the scaling
of the P-SFS flux in Figure~\ref{fig:pi_sq_cbH1} 
using a sharp Galerkin projector, which agrees with
the scaling in Equation~\eqref{eq:interm_scaling}, even though the rigorous upper bound that can be
obtained formally diverges for such a filter. This indicates that the upper
bound becomes less useful (less tight) even though the P-SFS flux still scales
as the F-SFS, \obs{that uses} 
a smooth filter.
\\ 

\noindent
The intermittent scaling of the SFS energy transfer was first investigated in {\em a-priori} as 
well as {\em a-posteriori} analyses in Ref. \cite{Cerutti98} at moderate Reynolds \obs{numbers} 
using \obs{both} a Gaussian filter and a sharp cut-off in Fourier space, but without applying the double filtering proposed here (\ref{eq:tau_ples}).  
By using  ESS \cite{Benzi93,Benzi95}, it was shown in \cite{Cerutti98} that 
the scaling of the SFS energy transfer is slightly 
more intermittent than the LSFs while being less intermittent
than the TSFs. However, as pointed out in Ref.~\cite{Cerutti98}, 
the accuracy of the measurements was not sufficient to 
warrant interpretation of the small differences in the exponents. \\
In what follows, we intend to perform a similar analysis at much higher Reynolds numbers, such as to avoid the use of ESS, 
and by comparing different filtering strategies and by analysing different components of the SFS energy transfer. \\
 \noindent
Figures \ref{fig:pi_sq_cbH1} \obs{(a-c)} presents the scaling of the SFS energy transfer 
compared to the predictions from Equation~\eqref{eq:interm_scaling}
for the different components 
of the P-SFS energy transfer obtained through Galerkin truncation.  
and for the F-SFS energy transfer. In all figures we superpose  
the multifractal prediction using either the longitudinal or the transverse scaling. Indeed, the SFS energy transfer  is a scalar quantity that cannot distinguish among the two scaling\obs{s} 
and one has to interpret the mismatch \obs{between}  
the two set\obs{s} of exponents as an estimate of the error \obs{i}n the scaling properties of mixed 
observables. 
From \obs{Figure~\ref{fig:pi_sq_cbH1}} 
it can be seen that the scaling of all 
components 
is consistent with Equation~\eqref{eq:interm_scaling} for $n=2,3$ and $n=4$.
The Leonard component $\langle (\Pileo)^n\rangle$ also scales in agreement with Equation~\eqref{eq:interm_scaling}
for the even orders $n=2,4$ as can be seen in Figures \ref{fig:pi_sq_cbH1} (a) and (c). The odd order $n=3$
is in very small agreement with the symmetry in the pdf already discussed. 

\begin{figure}[h]
\begin{center}
  \includegraphics[scale = 0.44]{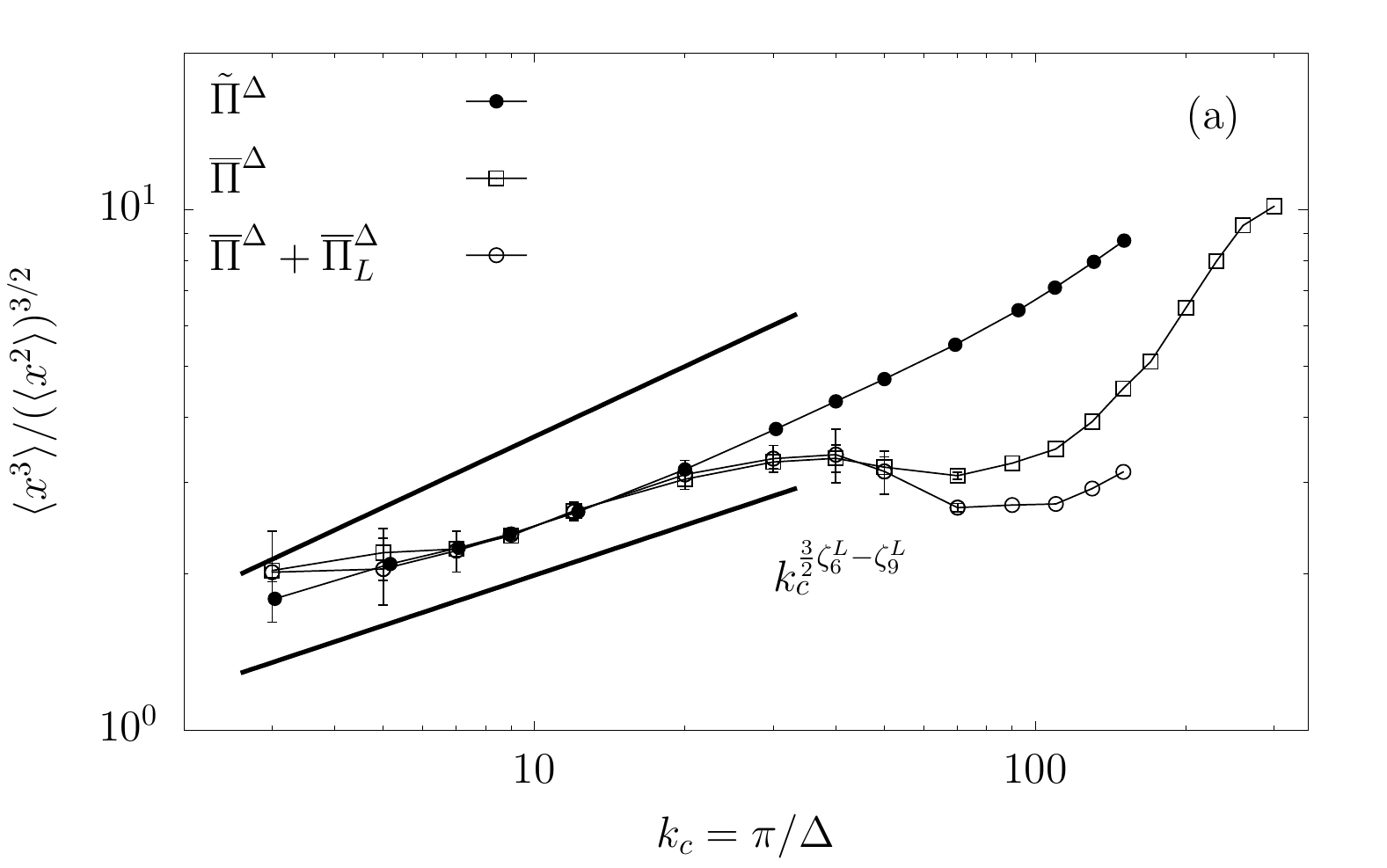}
  \includegraphics[scale = 0.44]{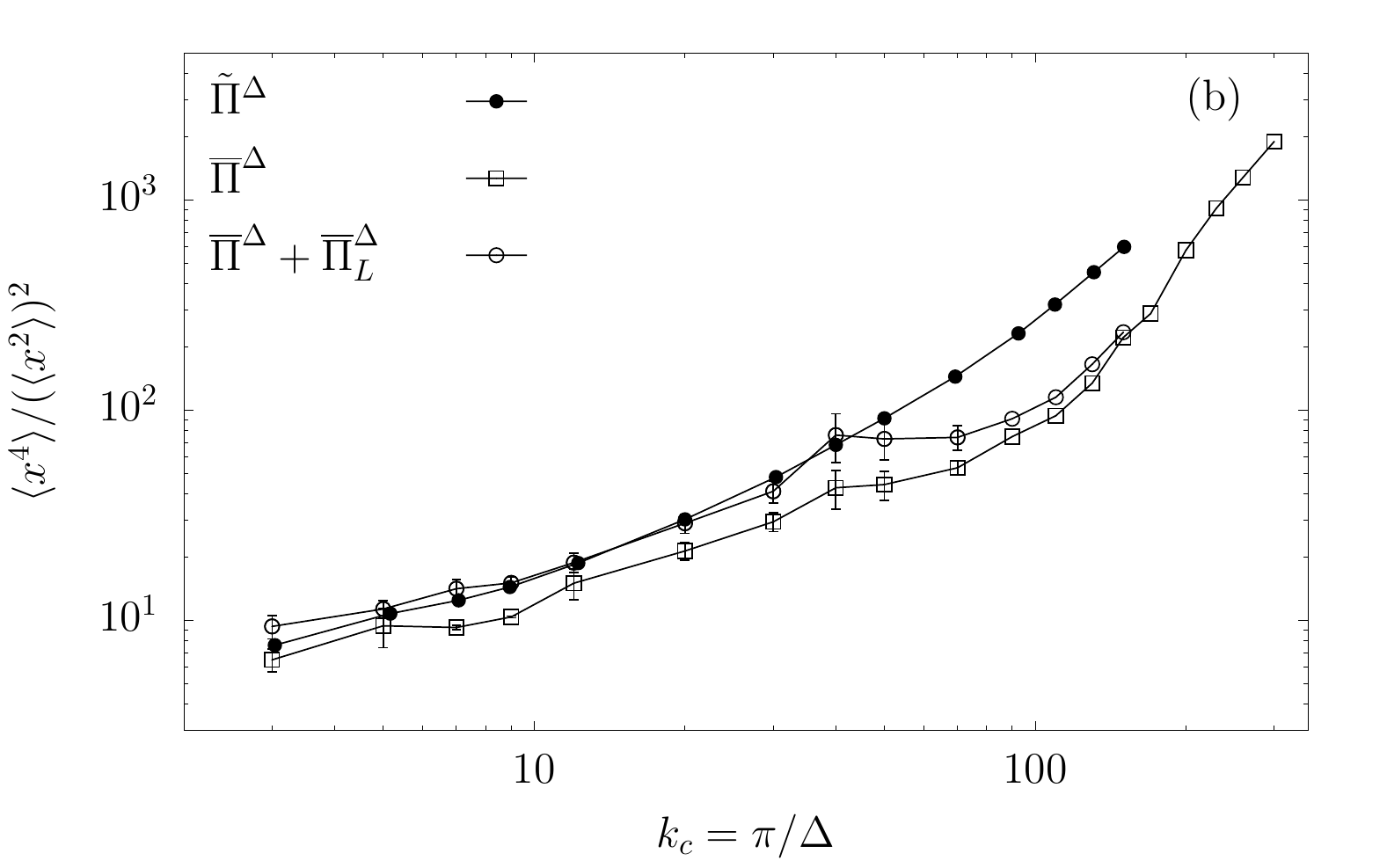}
  \caption{
             Skewness (a) and flatness (b) of the
             different components of P-SFS and F-SFS
             energy transfer for data-set H1.
          The solid lines in (a) indicate the scaling expected from the
          multifractal model and Equation~\eqref{eq:interm_scaling}
          using the anomalous exponents for the
          longitudinal and transverse structure functions
          $\zeta_{3n}^L$ and $\zeta_{3n}^T$, Ref.~\cite{Gotoh02}.
          The data corresponding to $\oPi+\Pileo$ and $\tPi$ have been
          shifted upward and downward, respectively, for presentational reasons.
}
\label{fig:pi_skew_flatH1}
\end{center}
\end{figure}
\noindent

Figure \ref{fig:pi_skew_flatH1} shows the skewness and flatness of the 
P-SFS energy transfer $\oPi$, its combination with the energy transfer due to the 
Leonard stresses $\oPi + \Pileo$
and the F-SFS energy transfer $\tPi$ obtained through Gaussian filtering.
Both skewness and flatness show similar scaling 
for $\oPi$, $\oPi + \Pileo$ and $\tPi$, however, the scaling range of  
the skewness corresponding to the F-SFS $\tPi$ has a smoother transition when crossing the viscous scales
compared to that 
obtained by sharp Galerkin projection. This is an indication of the importance of contributions
 from a wide range of scales affecting the F-LES formalism.   
In other words, since the smooth Gaussian filter is less localised 
in $k$-space compared to the sharp projector, 
it retains contributions from a larger number of Fourier modes at different
wavenumbers, see also \cite{Rivera14} for an illustration.
Concerning the flatness, there
appears to be little difference between $\oPi + \Pileo$ and $\oPi$; the two corresponding
curves for the flatness nearly collapse in the intermittent scaling range without any shift in the 
data.  
\\
\noindent 
In summary, the second, third and fourth-order moments as well as the skewness and flatness
of the SFS energy transfer for the sharp filter show intermittent scaling for both 
$\oPi + \Pileo$ and for the P-SFS definition $\oPi$ alone, in agreement with the local-estimates based on the bridge relation among 
the SFS energy transfer at filter width $\Delta$ and velocity increments at scale $r \sim \Delta$. 
\subsubsection{Reynolds \obs{number} dependency}
\begin{figure}[H]
\begin{center}
  \includegraphics[scale = 0.7]{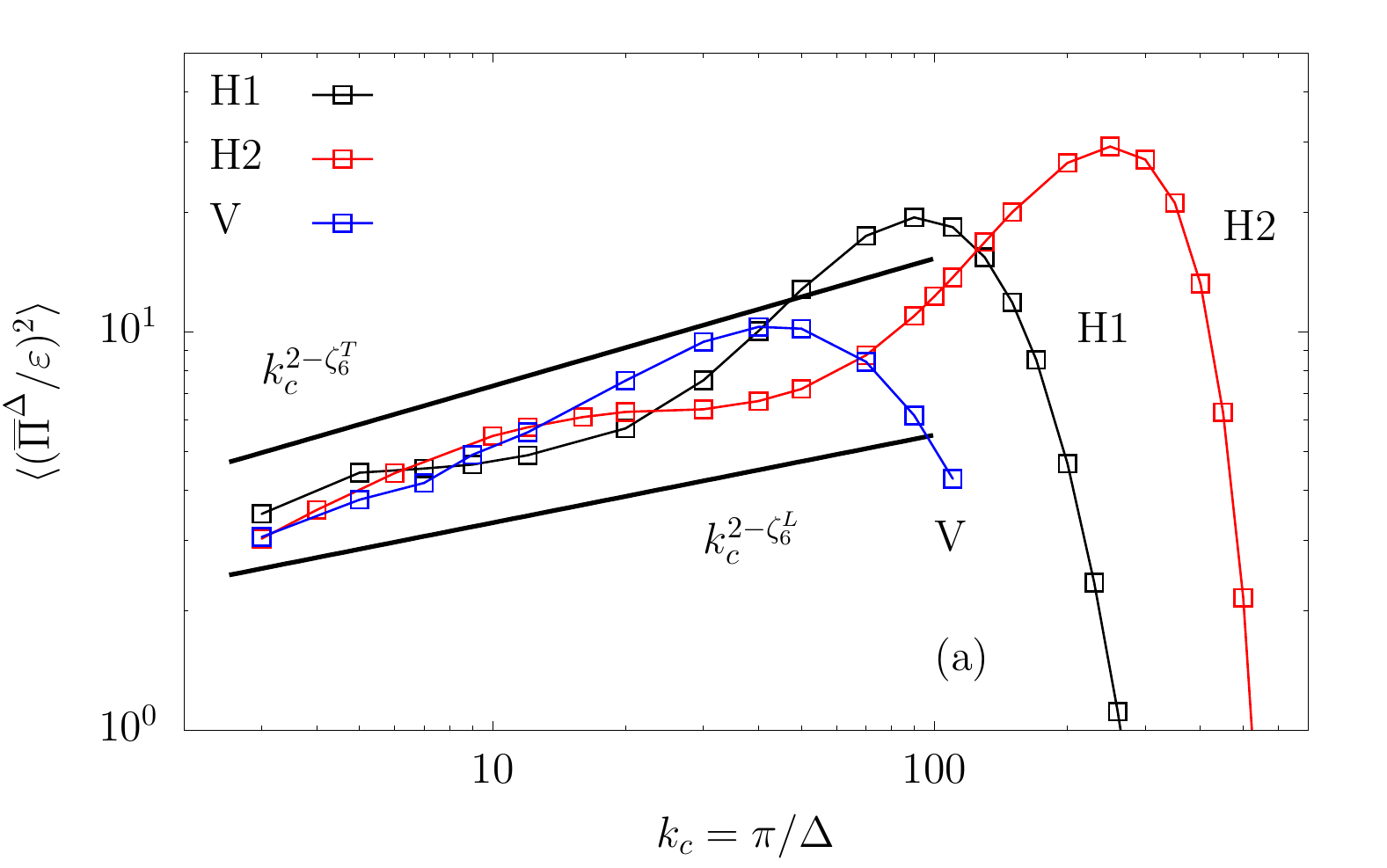}\\ 
  \includegraphics[scale = 0.7]{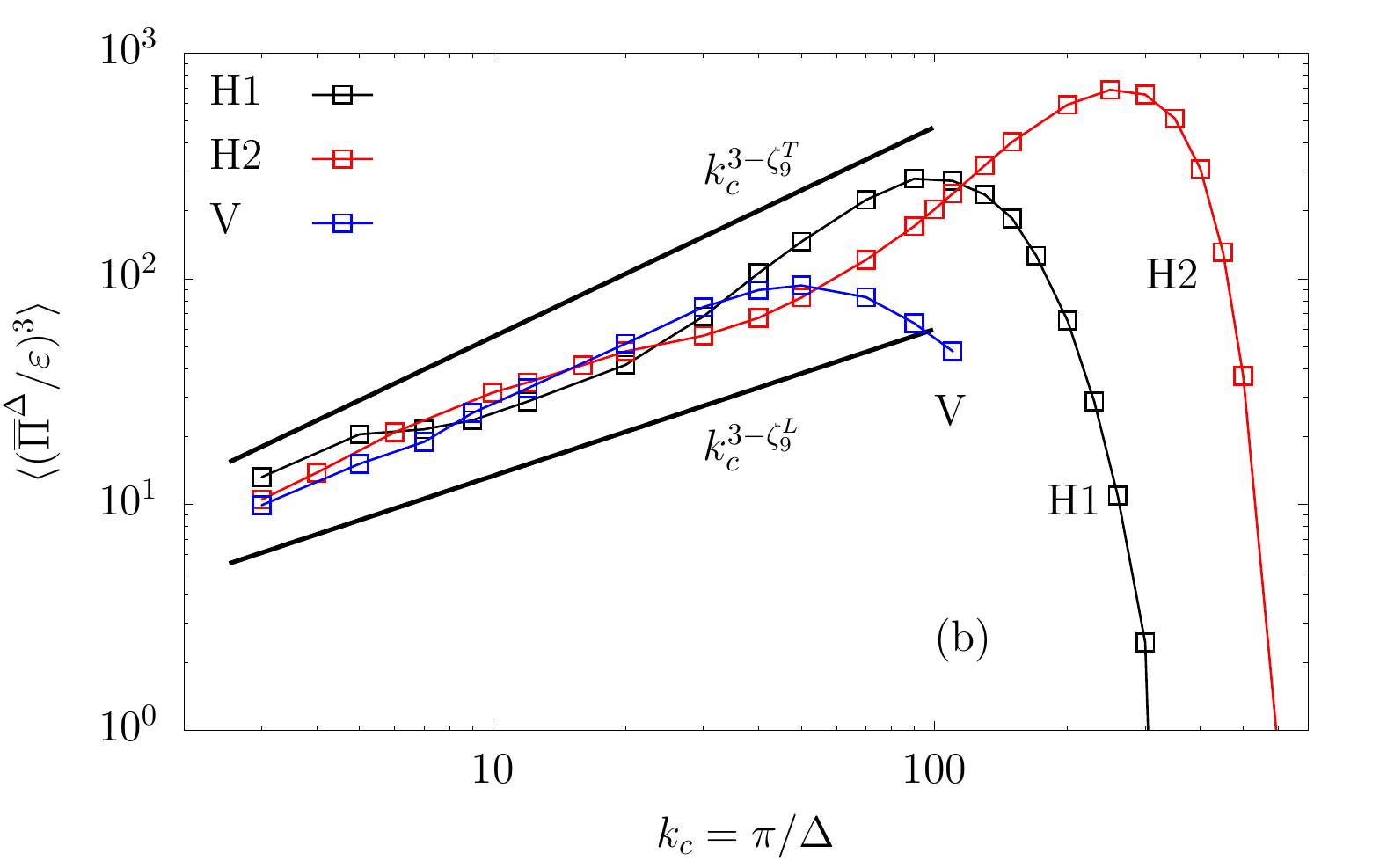}\\ 
  \includegraphics[scale = 0.7]{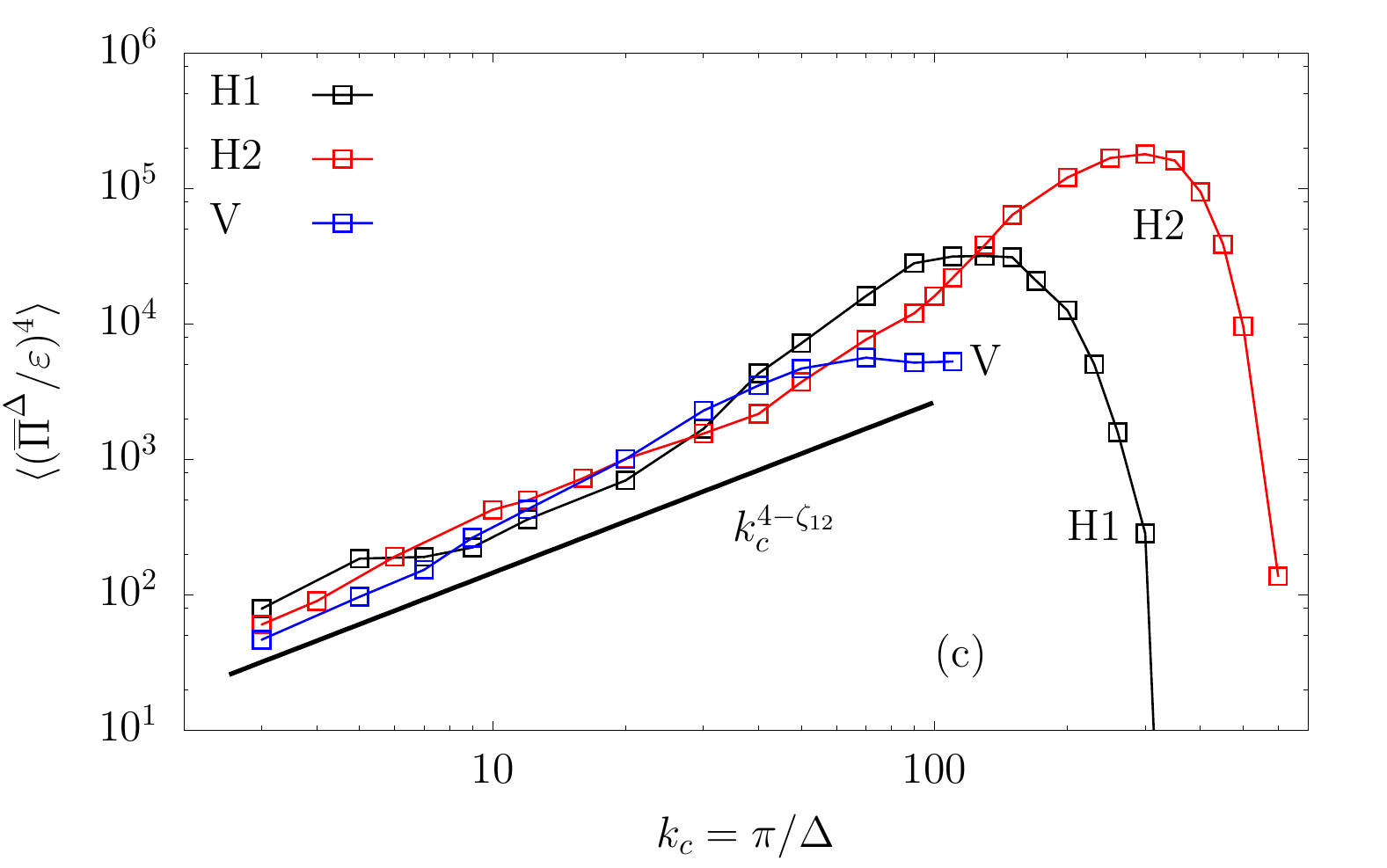}
  \caption{
             \mb{Scaling of the $n^{\textrm{th}}$ moments of
             the normalised P-SFS energy transfer for data-sets V1 (blue/dark gray), V2 (green/light gray), H1 (black) and H2 (red/gray) 
             as a function of $k_c = \lb{\pi}/\Delta$,
         (a) $n=2$,
         (b) $n=3$,
         (c) $n=4$.
          The solid lines indicate the scaling expected from the
          multifractal model and Equation~\eqref{eq:interm_scaling}
          using the anomalous exponents for the
          longitudinal and transverse structure functions $\zeta_{3n}^L$ and $\zeta_{3n}^T$, Ref.~\cite{Gotoh02}.
          In (c) for $n=4$ the prediction from the She-L\'ev\^eque model $\zeta_{3n} = 2.74$ \cite{She94,Boffetta08} is shown.}}
\label{fig:pi_sq_cb}
\end{center}
\end{figure}

In the remainder of this section, we assess the results for $\oPi$ by 
extending the analysis to both a larger and smaller inertial range using the three additional data-sets V1, V2 and H2 described in table \ref{tbl:simulations}. Results for the scaling of $\langle (\oPi)^n \rangle$, 
obtained from data-sets H1, H2  and V1, V2 are presented in Figures~\ref{fig:pi_sq_cb}(a-c) for $n=2,3$ and $n=4$, respectively. In Figure~\ref{fig:pi_sq_cb}(a) we observe that the intermittent 
scaling extends to a larger range of scales for the higher Reynolds number data-set H2 and to a 
shorter range  for data-set V1. We note that the effect of the bottleneck \cite{Falkovich94} in hyperviscous simulations is visible in the statistics of the SFS energy transfer. As can be seen 
in Figures~\ref{fig:pi_sq_cb}(b,c), the two hyperviscous simulations H1 and H2 consistently display 
a much larger deviation from intermittent scaling towards the end of the inertial range
compared to the Newtonian viscous simulations V1 and V2. 
 In other words, scaling-wise there is not such a big
gain by moving from normal to hyperviscosity (compare data sets V and H),
unlike for the extension of the range where the total energy flux is constant 
(see Figure~\ref{fig:pi_normalised}). 
Similar observations have been made in Refs.~\cite{Frisch08,aluie2009I,aluie2009II}.

\section{Comparison of different projector filters} \label{sec:filters}
\begin{figure}[H]
\begin{center}
  \includegraphics[scale = 0.44]{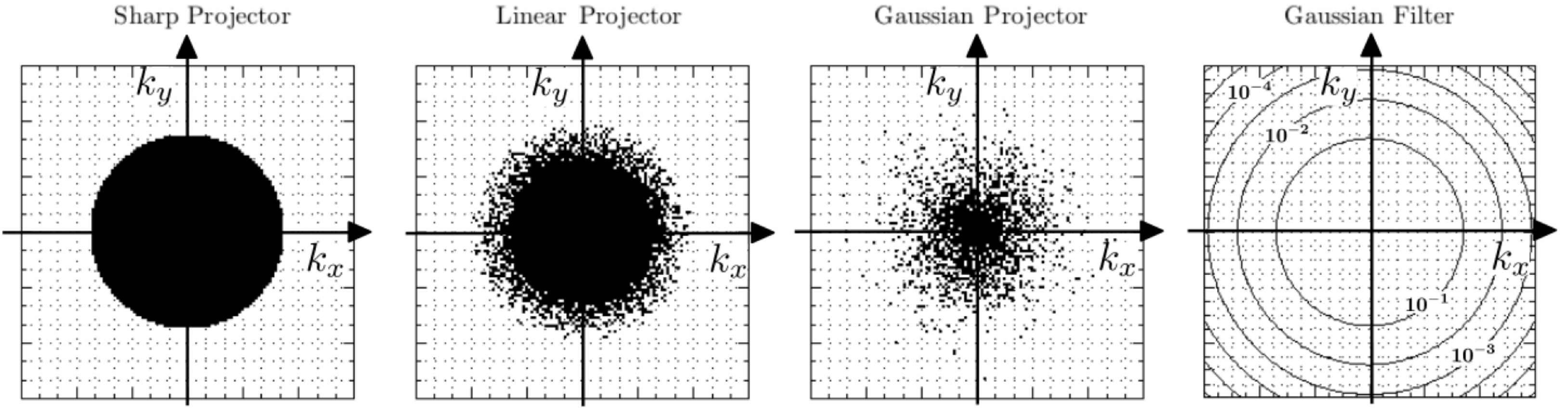}
  \caption{
Two-dimensional illustration of different filters in Fourier space. 
From left to right: 
 sharp Galerkin projector above $k_c$,
 Galerkin projector with a linear profile (in probability) between $k_c$ and  $2k_c$,
 Galerkin projector with a Gaussian profile (in probability),
 Gaussian filter.
}
\label{fig:illustration}
\end{center}
\end{figure}

As \obs{pointed out} 
in the Introduction, different filters 
introduce different 
fluctuations and the two traditional \obs{filters,} 
Gaussian smoothing and sharp Galerkin projection, 
each have their own limitations. 
Although the Gaussian filter results in a positive definite SFS \obs{stress} tensor \cite{Vreman94a} 
and does not induce Gibbs oscillations, it has the important limitation of 
not producing a clear distinction between resolved and unresolved scales.
In an attempt to improve on the drawbacks of both 
traditional 
approaches of sharp Galerkin truncation and 
Gaussian smoothing, we introduce a new family of projector filters for which the 
truncation operation is carried out in a probabilistic way \cite{Frisch12}. 
\obs{Specifically,} 
the truncated modes are chosen randomly according to a 
probability density which decreases either linearly with increasing wavenumber:
\be
\label{eq:lin_proj}
\hat G_\Delta(\bk) =
\begin{cases}
1 \quad \mbox{for} \ \ |\bk| < k_c \\
1 \quad \mbox{with probability} \ \ P(k) = \frac{\lambda -k/k_c}{\lambda -1} \ \ \mbox{for} \ \ k_c < |\bk| \leqslant \lambda k_c \\
0 \quad \mbox{for} \ \ |\bk| > \lambda k_c \ ,
\end{cases}
\ee
where $k_c = \lb{\pi}/\Delta$ and $\lambda > 1$,
or according to a Gaussian probability density as
\be
\label{eq:gauss_proj}
\hat G_\Delta(\bk) =
1 \quad \mbox{with probability} \ \ P(k) = e^{-k^2\Delta^2} \ . 
\ee
Two-dimensional graphical representations of all filters in Fourier space
are shown in Figure~\ref{fig:illustration}. 
\\
\noindent
The SFS energy transfer obtained through the linear probabilistic projector acting at a 
given threshold $k_c$ leads to a  `band-averaged' Fourier-space flux
\be
\label{eq:band_flux}
\langle \oPi \rangle = \frac{1}{(\lambda-1)k_c}\int_{k_c}^{\lambda k_c} dk \ \Pi(k) \ . 
\ee
The `band-averaged' Fourier flux had been introduced in Ref. \cite{eyink94} 
in order to study the locality of triadic interactions and can be
obtained through a filter $G_\Delta$ whose Fourier-space profile $\hat G_\Delta^2(k)$ 
is a linearly decreasing function of $k$ \cite{eyink2}. The introduction of 
 (\ref{eq:lin_proj}) must be seen as a way to reproduce the same spectral properties of the filter proposed
in  \cite{eyink2} but with the added feature  of being a projector. \\
The proof of Equation~\eqref{eq:band_flux} for the linear projector follows 
\obs{from} minor modifications \obs{of}
the corresponding proof of the linear filter in Ref.~\cite{eyink2}.
In the remainder of this work, the linear projector is applied with $\lambda=2$
unless specified otherwise.
\\

\begin{figure}[H]
\begin{center}
  \includegraphics[scale = 0.44]{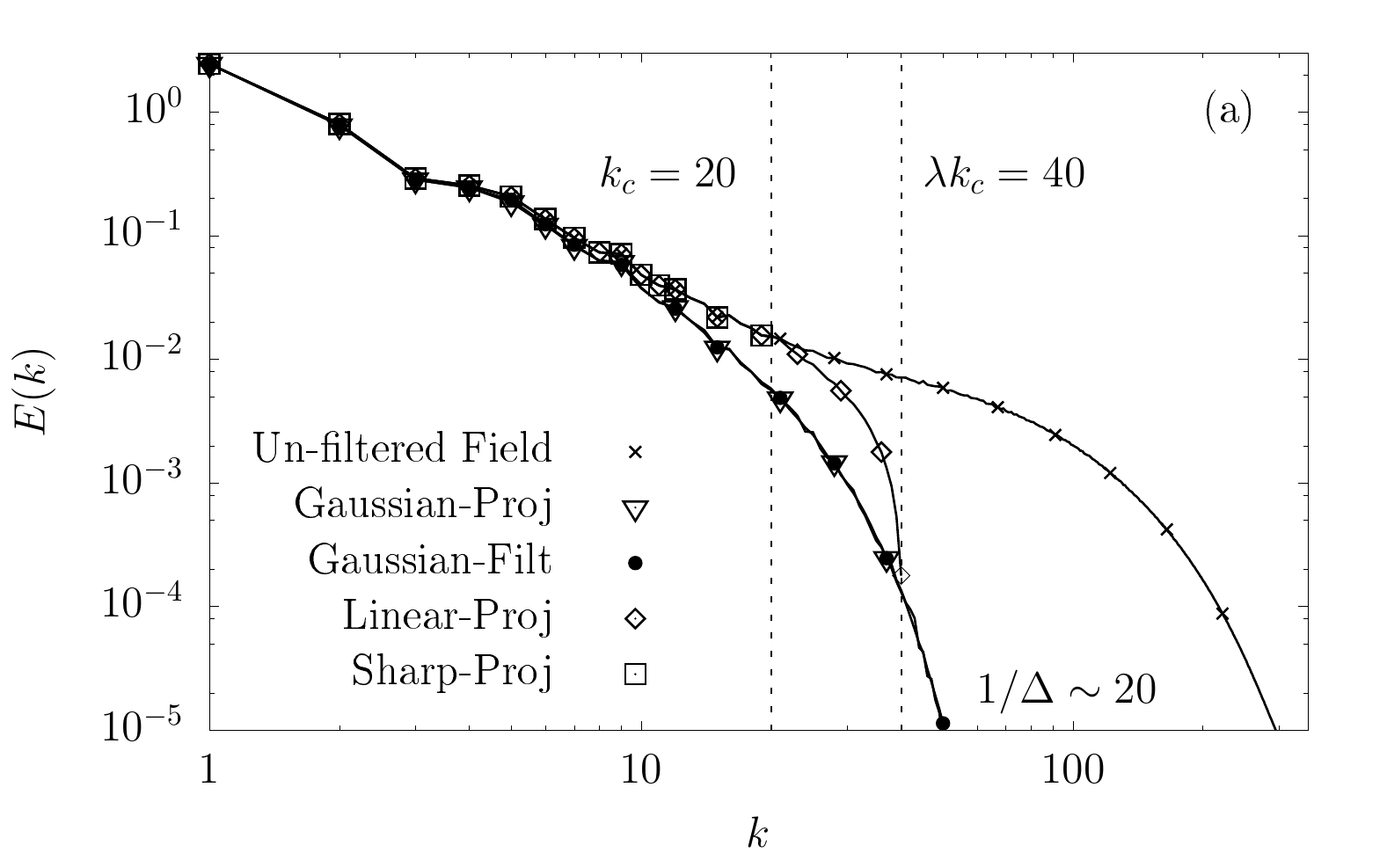}
  \includegraphics[scale = 0.44]{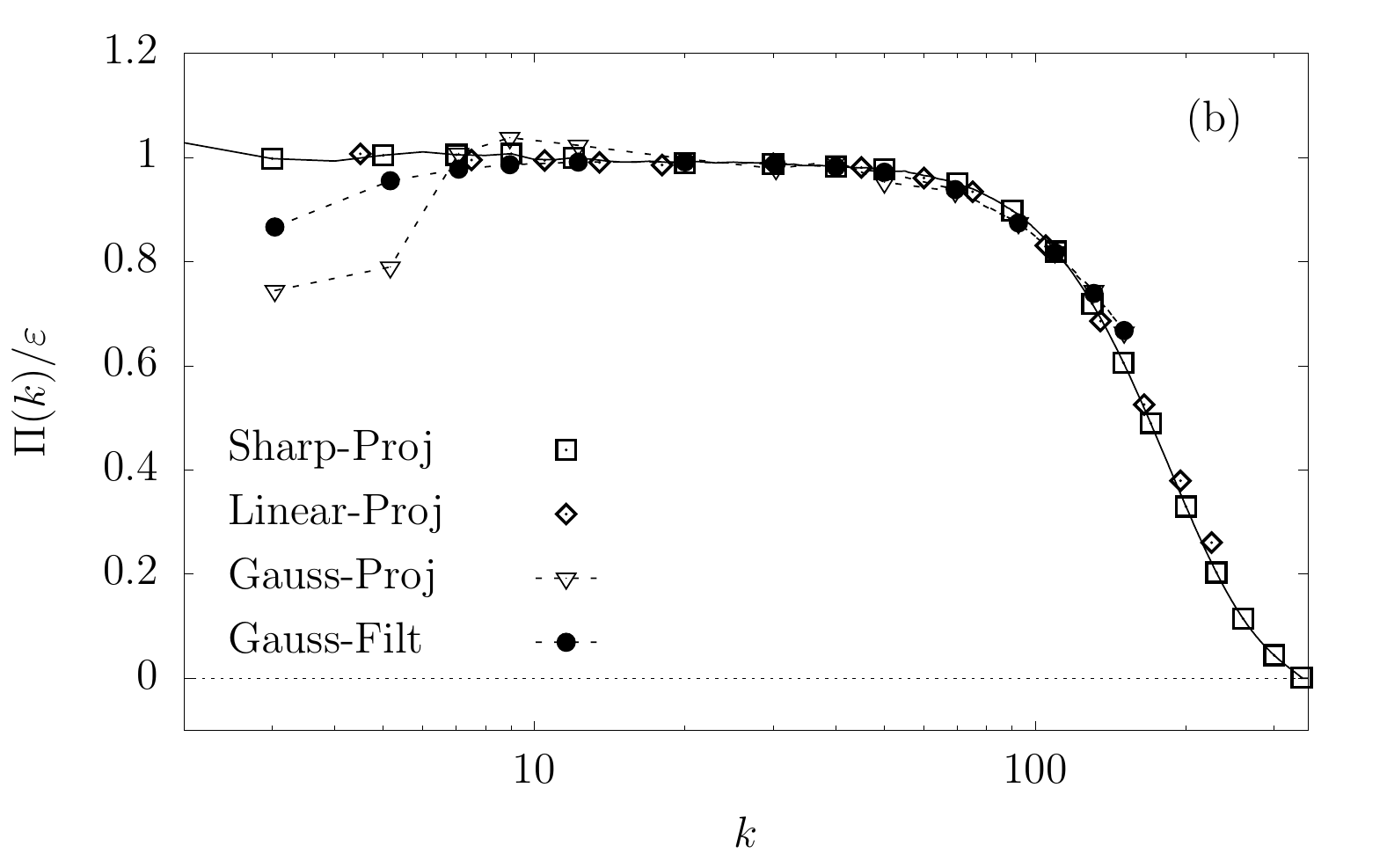}
  \caption{
 data-set H1. (a) Energy spectra obtained by different filtering procedures.
 (b)  Comparison between the Fourier-space energy flux and the SFS energy transfer
  as a function of the cut-off wavenumber $k_c$ for the sharp projector (squares), as a function of 
  $3k_{c}/2$ for the linear projector (diamonds) and
  as a function of $\lb{\pi}/\Delta$ for the Gaussian projector (triangles). The SFS energy transfer obtained through Gaussian smoothing is shown for comparison (black dots).
}
\label{fig:pi_normalised_proj}
\end{center}
\end{figure}

\noindent
We now proceed to investigate the new projectors in comparison with the sharp
Galerkin projector using only the data-set H1,
where for all projectors we consider the SFS-tensor corresponding to the P-SFS approach.
The energy spectra obtained by different filtering procedures are presented in 
Figure~\ref{fig:pi_normalised_proj}(a) alongside the original unfiltered data. 
As can be seen, the linear projector results in a smooth roll-off of the spectrum 
between $k_c$ and $2k_c$. The energy spectra obtained through Gaussian filtering and 
probabilistic Gaussian projection are indistinguishable, as expected.  
The mean SFS energy transfers 
obtained using the different projectors
are shown in Figure~\ref{fig:pi_normalised_proj}(b) 
as functions of $k_c = \lb{\pi}/\Delta$, in comparison with the Fourier-space energy flux $\Pi(k)$. 
The mean SFS energy transfer obtained from the linear projector 
agrees well with $\Pi(k)$ in the inertial range, as expected. The agreement is still good 
in the beginning of the viscous range, where $\Pi(k)$ decreases linearly and thus should 
coincide with $\langle \oPi \rangle$ 
obtained using the linear projector if plotted against $k=3k_c/2$ as is the case in 
Figure~\ref{fig:pi_normalised_proj}(b). 
Deviations between the two fluxes become visible only at relatively high wavenumbers 
where the Fourier-flux $\Pi(k)$ decreases exponentially. 
The SFS energy transfer obtained from the Gaussian projector 
shows significant deviations from $\Pi(k)$ at low wavenumbers, 
while being in reasonable agreement in the inertial range. 
Proceeding towards the viscous range, again we observe deviations from $\Pi(k)$. 
Similar to the deviations between the SFS energy transfers 
obtained by smooth Gaussian filtering, the deviations between $\Pi(k)$ and $\langle \oPi \rangle$ can indeed be 
expected for non-sharp projector filters. 
\\

\begin{figure}[H]
\begin{center}
  \includegraphics[scale = 0.7]{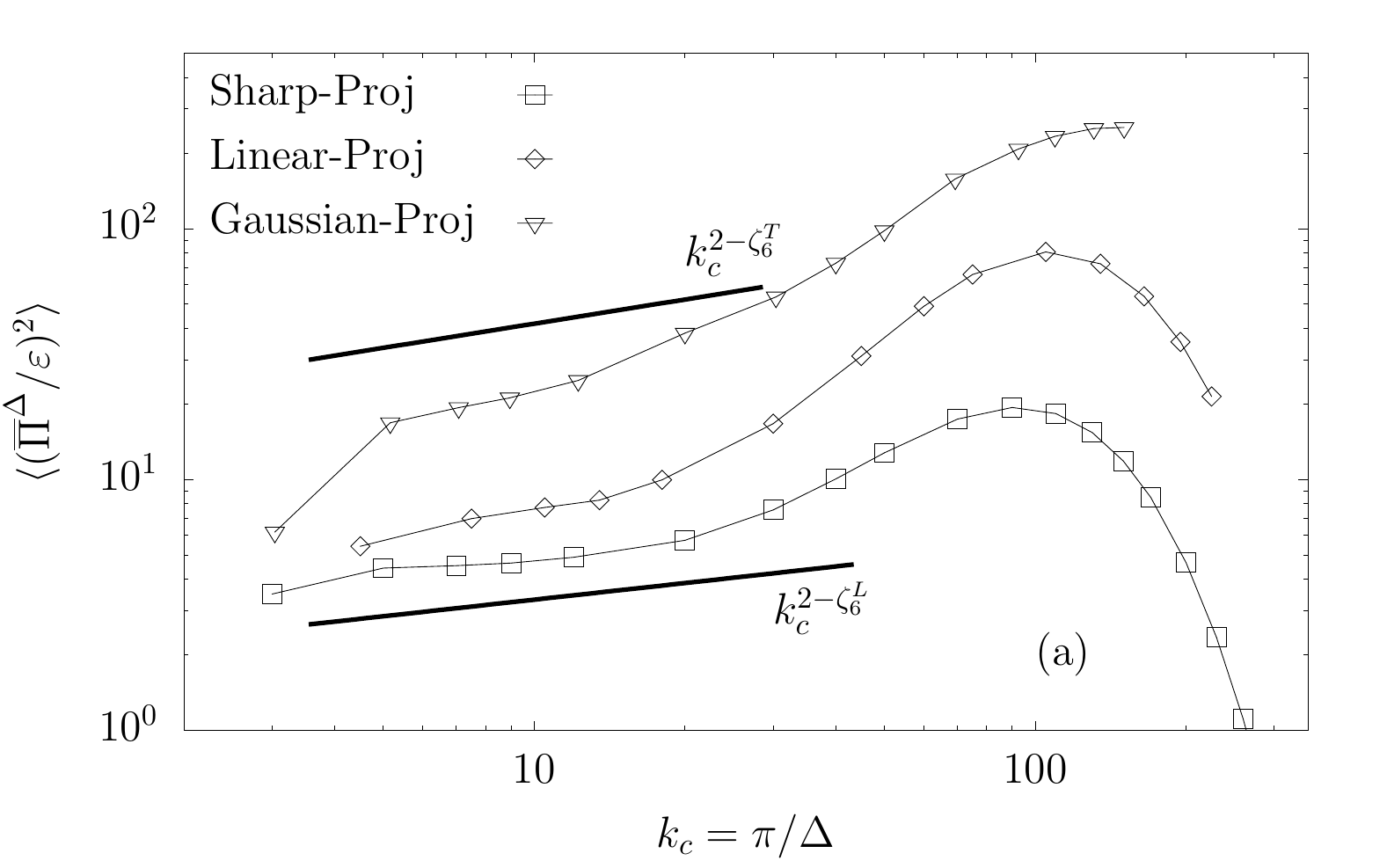} \\
  \includegraphics[scale = 0.7]{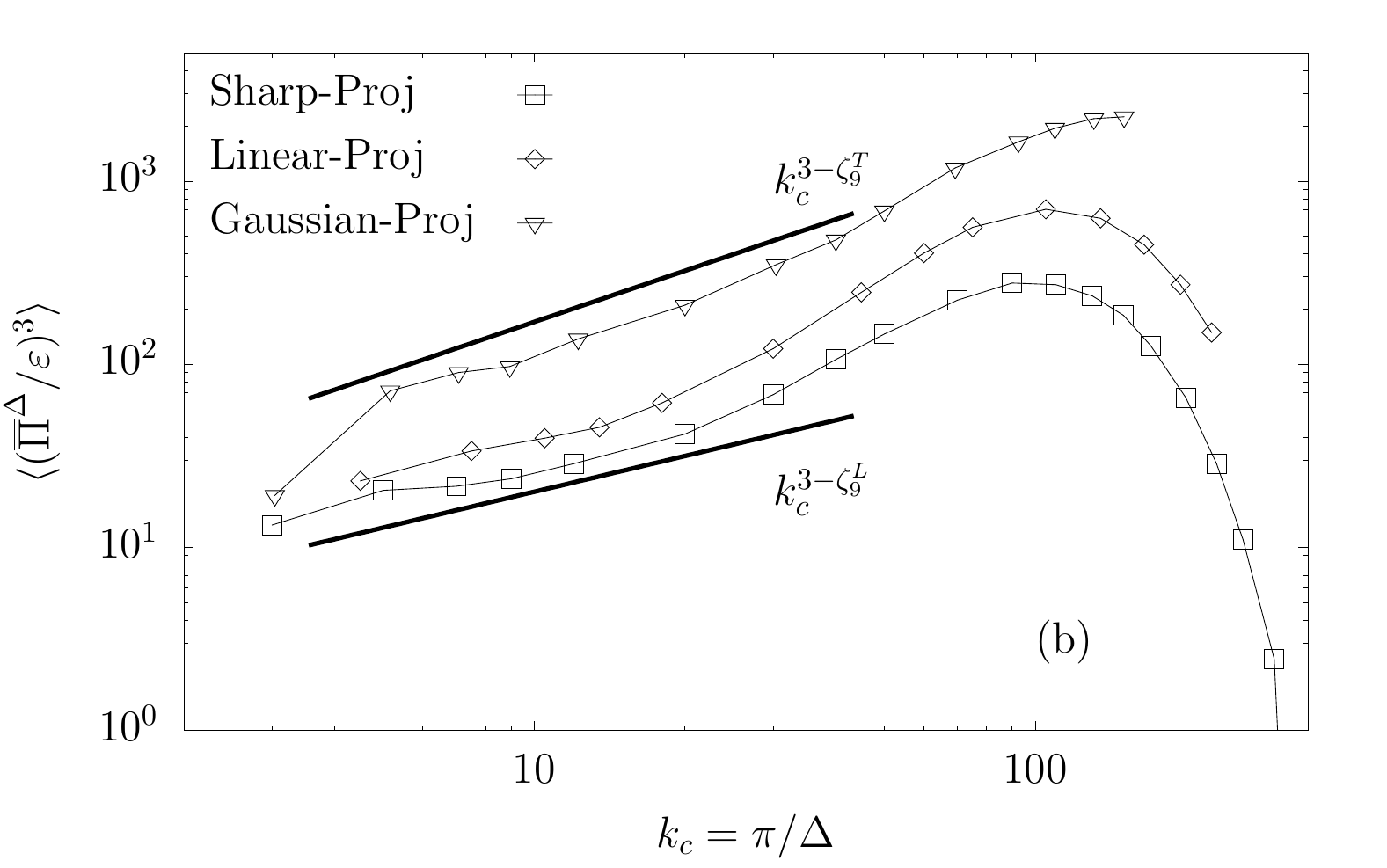} \\
  \includegraphics[scale = 0.7]{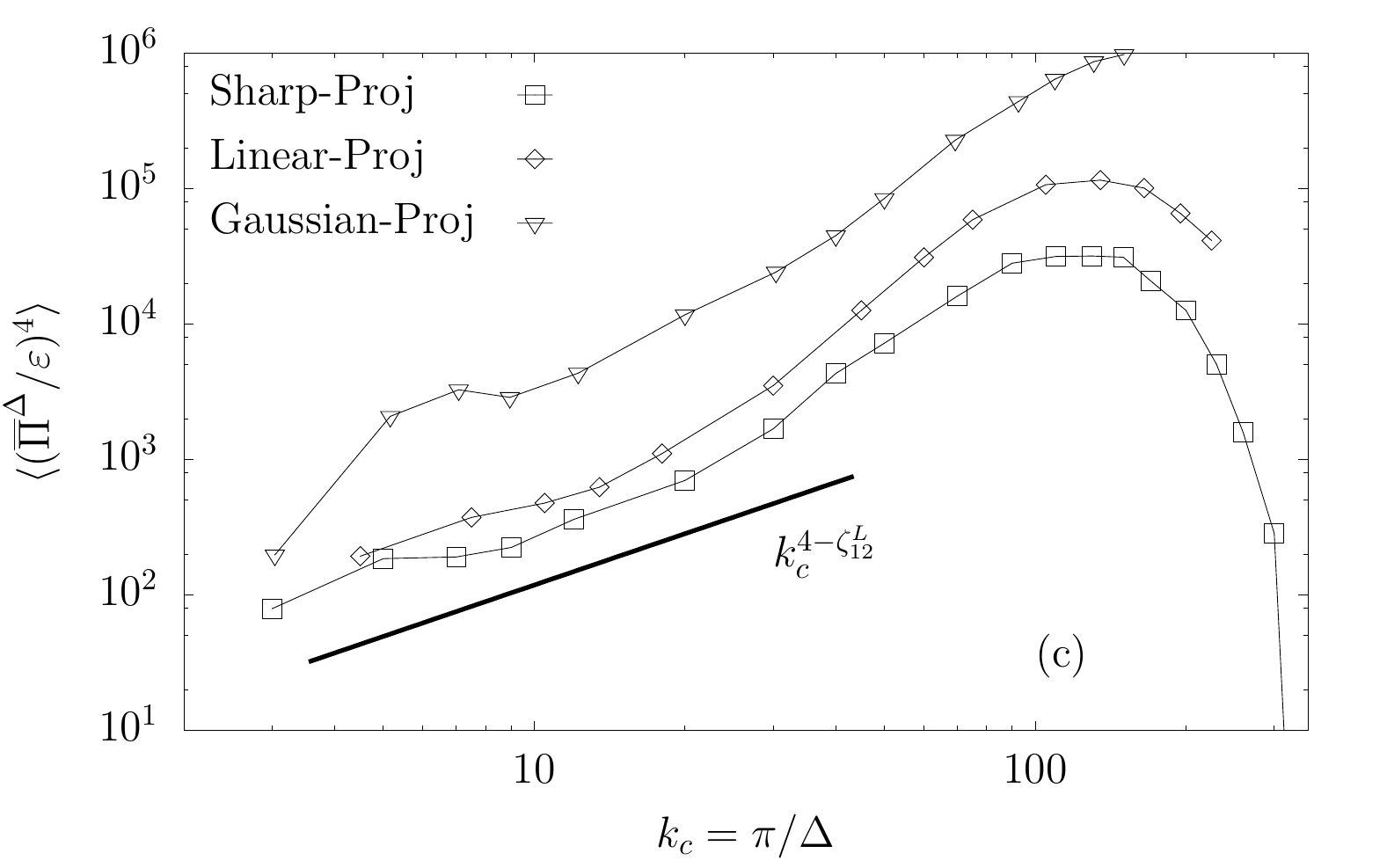}
  \caption{
             Scaling of the $n^{\textrm{th}}$ moments of the P-SFS energy transfer for the sharp projector 
             (squares) as a function of $k_c = \lb{\pi}/\Delta$, the linear projector
             (diamonds) as a function of $3k_c/2$
             and the Gaussian projector (triangles) as a function of $\lb{\pi}/\Delta$. Moments: (a) $n=2$,
         (b) $n=3$,
         (c) $n=4$.
          The solid lines indicate the scaling expected from the
          multifractal model and Equation~\eqref{eq:interm_scaling}
          using the anomalous exponents for the
          longitudinal and transverse structure functions $\zeta_{3n}^L$ and $\zeta_{3n}^T$, Ref.~\cite{Gotoh02}. In (c), for $n=4$, the solid line indicates the prediction from the She-L\'ev\^eque model $\zeta_{3n} = 2.74$ \cite{She94,Boffetta08} is shown.
}
\label{fig:pi_sq_projectors}
\end{center}
\end{figure}

\begin{figure}[H]
\begin{center}
  \includegraphics[scale = 0.44]{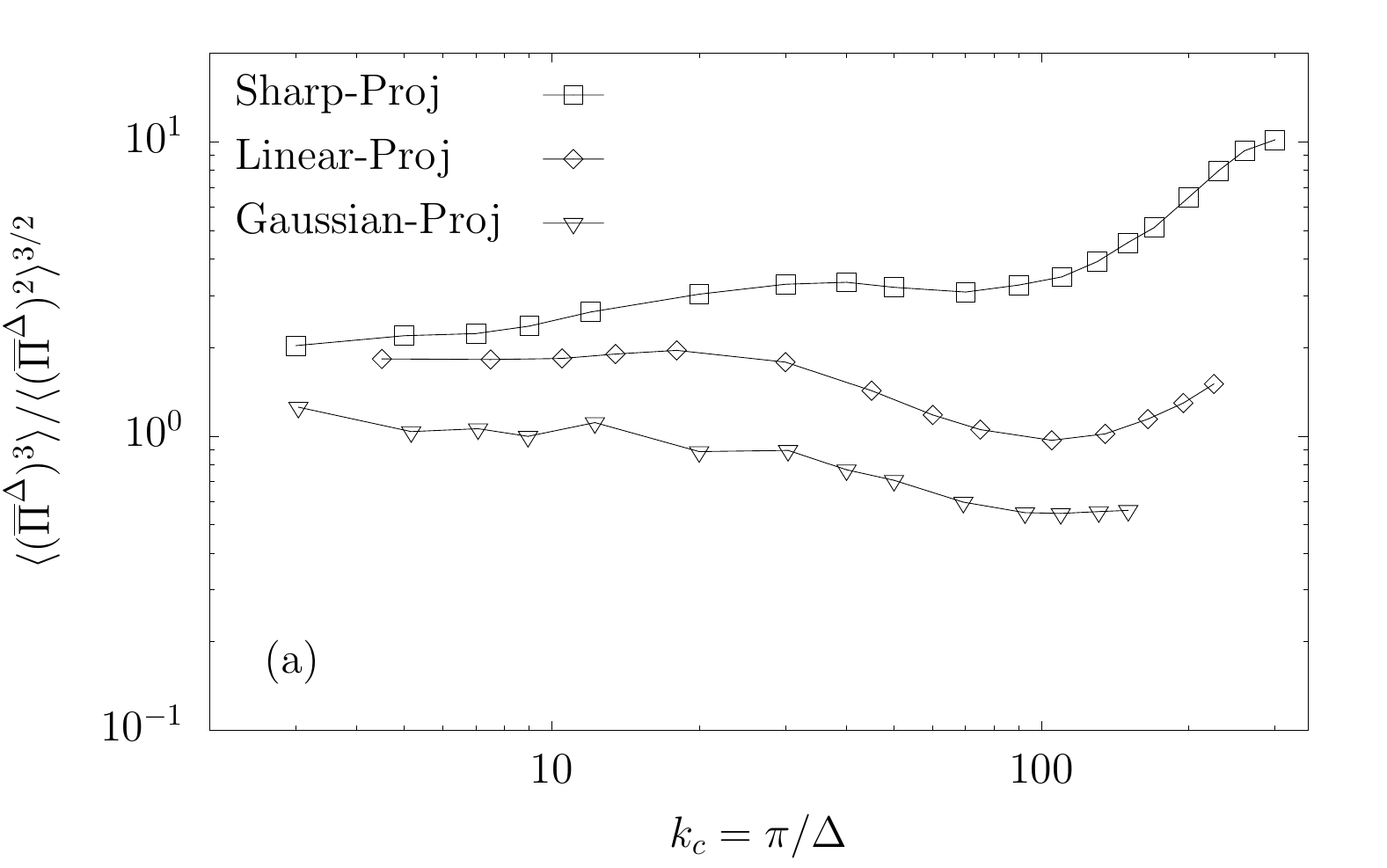}
  \includegraphics[scale = 0.44]{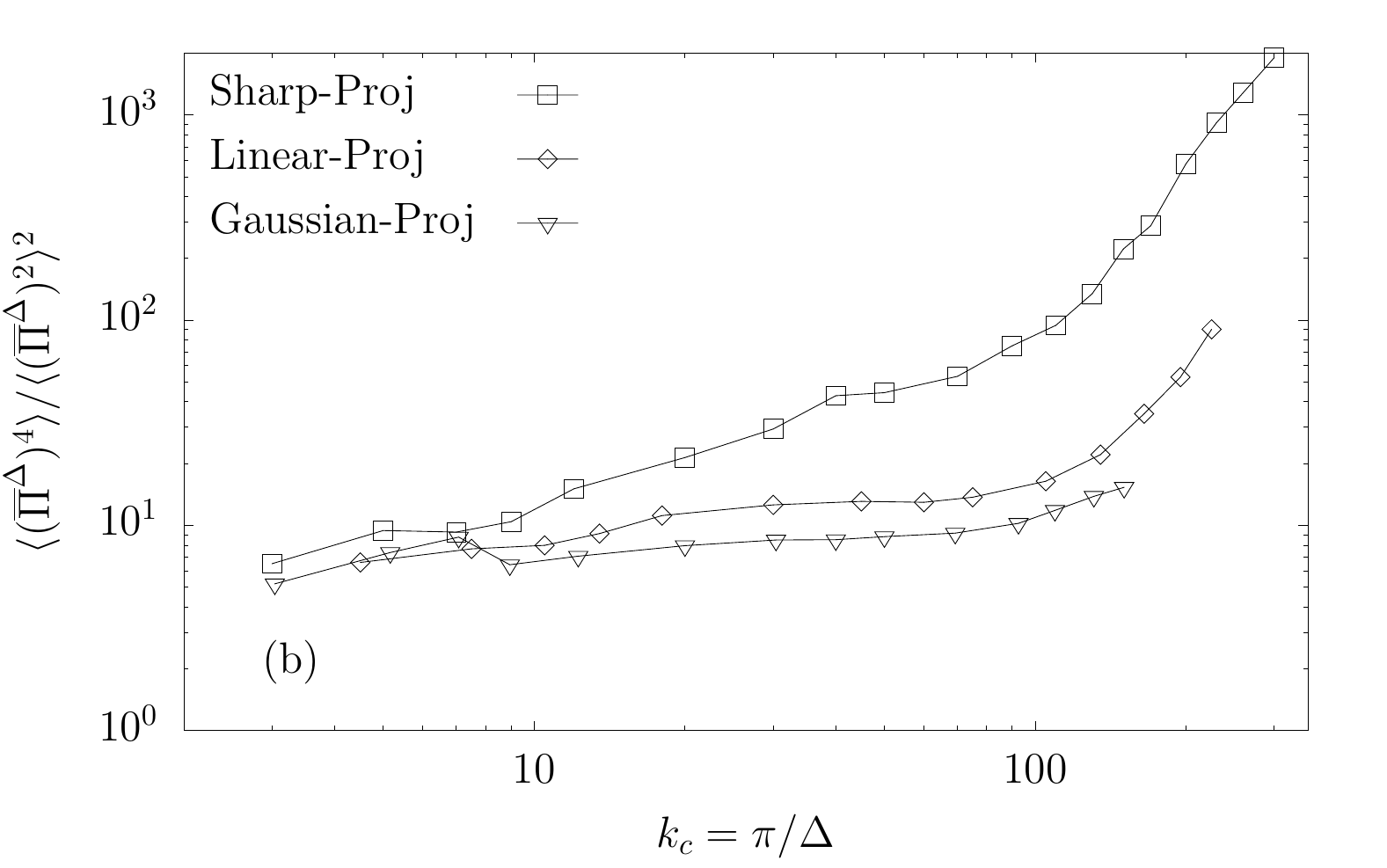}
  \caption{
 Skewness (a) and flatness (b) of the
             P-SFS energy transfer
         for the sharp (squares),
         the linear (diamonds) and the Gaussian (triangles) projectors for
         data-set H1.
          The solid lines in (a) indicate the scaling expected from the
          multifractal model and Equation~\eqref{eq:interm_scaling}
          using the anomalous exponents for the
          longitudinal and transverse structure functions
          $\zeta_{3n}^L$ and $\zeta_{3n}^T$, Ref.~\cite{Gotoh02}.
}
\label{fig:skew_flat_projectors}
\end{center}
\end{figure}
\noindent
Concerning the multiscale statistics of the P-SFS energy transfer $\oPi$ for the different projectors,
the non-sharp projectors scale differently compared to the sharp projector for the symmetric
part $\langle (\oPi)^2\rangle$ as shown in Figure~\ref{fig:pi_sq_projectors}(a), 
while all projectors display similar scaling for the asymmetric part 
$\langle (\oPi)^3\rangle$  and for $\langle (\oPi)^4\rangle$ as shown in 
Figures~\ref{fig:pi_sq_projectors} (b-c). 
In particular, the scaling of $\langle (\oPi)^2\rangle$ 
obtained using the Gaussian projector deviates significantly 
from the intermittent scaling displayed by $\langle (\oPi)^2\rangle$
obtained by sharp Galerkin projection. 
The difference in the scaling of $\langle (\oPi)^2\rangle$ 
between the different projectors propagates into the scaling of skewness and flatness as shown in 
Figures \ref{fig:skew_flat_projectors}(a,b)
\\
\noindent
The deviation from intermittent scaling of skewness and flatness of the 
SFS energy transfers for the linear and Gaussian projectors may be connected to their
higher degree of discontinuity in Fourier space. 
Previous studies \cite{Frisch12,Lanotte15,Buzzicotti16} showed that a removal of the 
degrees of freedom results in a decrease in intermittency. 
\obs{Such removal of the degrees of freedom can be carried out either by a dynamic
procedure where the corresponding projection operation is carried out at each iteration 
step, or by one-off projection carried out on the DNS data obtained by evolving the full 
Navier Stokes equations. The latter is referred to as static decimation.}
\obs{The} 
dynamic fractal projection operation generally leads to a drastic decrease 
in intermittency, where the removal of a small percentage of Fourier modes already results in near-Gaussian statistics at all scales. 
Intermittency is also decreased by static decimation \cite{Lanotte15}, however, compared to the 
dynamic procedure a much larger percentage of modes must be removed. 
The probabilistic filtering applied here could be seen as a static decimation carried out in 
logarithmically spaced Fourier bands, where at least towards the middle of the wavenumber band a significant 
percentage of Fourier modes will have been removed. 
A non-sharp projector has also the disadvantage of increasing the 
frontier in Fourier space among resolved and unresolved modes
and it induces further discontinuities in Fourier space through the probabilistic projection
operation. 
As such, Gibbs oscillations in real space are enhanced, 
perhaps 
one of the reasons \obs{for the reduction in intermittency.} 
These properties might not 
necessarily 
\obs{be} detrimental for the implementation of {\it a posteriori} SFS stress models
on such a non-traditional Fourier support. The effects of the SFS stress tensor on the dynamical evolution of the resolved scale can be summarised in 
a multiscale correlation function among velocity increments $\delta_r \bv$ and the SFS tensor $\tau(\Delta)$ with $r >\Delta$ (see 
\cite{Meneveau94} for a discussion concerning the evolution of second order correlation functions of the resolved field),  
and it remains  to be checked by explicit LES performed on different 
Fourier supports how to minimise the feedback \obs{in order to} 
enhance the extension 
of the inertial range. \obs{W}ork in this direction will be reported elsewhere. 
\section{Conclusions} \label{sec:conclusions}
\noindent
In this paper, we \obs{have} studied the statistical properties of SFS energy
transfer in the inertial range of scales with particular emphasis on 
the effect of the filtering procedure using  approaches 
based on  Gaussian smoothing, sharp Galerkin projections and  new 
multiscale projectors.
We discuss formal and  practical advantages/disadvantages of projector filters and 
we discuss a LES formalism which is  Galilean invariant and mathematically well defined. 
In order to assess the multiscale statistics and the related scaling estimates,
we carried out an {\em a priori} analysis of the SFS energy transfer 
obtained through different filtering procedures using  
high-resolution DNS data-sets with normal as well as hyperviscosity 
on up to $2048^3$ collocation points.
We extend known results for the scaling properties of the SFS energy transfer using Gaussian smoothing \cite{eyink2,Eyink05}
to the case of sharp projector filters relating the
scaling exponents of the SFS energy transfer to the anomalous exponents 
of the velocity structure functions.         
\noindent
We find that the SFS energy transfer is sensitive 
to intermittent effects. 
Although the intermittent scaling of SFS energy transfer  appears to be
sensitive to the additional oscillations induced by the probabilistic
projectors, 
the effects of different filtering protocols on \emph{a posteriori} LES simulations \lb{remain to be studied}.
 It may even be conceivable
that a filter which induces more oscillations in physical space results in a
decorrelation effect between the resolved scales and the scales close to the
filter threshold which are most affected by the choice of filtering strategy.
Our results 
\lb{can be regarded as a systematic assessment of the impacts
of using projectors or filters on}
the multiscale properties of turbulence at high Reynolds numbers and
prompt for the need to perform suitable LES {\em a posteriori} studies to benchmark
the validity of different subgrid models to reproduce those properties.
\noindent
\section*{Acknowledgements}
The research leading to these results has received funding from the European
Union's Seventh Framework Programme (FP7/2007-2013) under grant agreement No.
339032 and from COST ACTION  (MP 1305). 
H. Aluie was also supported through NSF grant OCE-1259794, DOE grants DE-SC0014318 
and DE-561 NA0001944, and the LANL LDRD program through project 
number 20150568ER.
J. Brasseur was supported by AFOSR Grant FA9550-16-0388. 
C. Meneveau was supported by the National Science Foundation (CBET 1507469).

\appendix
\section{Galilean invariance}
\label{app:GalInv}
We consider a Galilean transformation
\be
\label{eq:galilean-transf1}
x_i \rightarrow x_i - u_i^0t  \ ,
\ee
such that
\be
\label{eq:galilean-transf2}
v_i \rightarrow v_i + u_i^0 \ ,
\ee
with a spatially uniform and time-independent $u_i^0$.
The aim is to establish  
(i) the breaking of Galilean invariance of the 
SFS stress tensor induced by additional filtering of the 
inertial term by a non-projector filter, and 
(ii) the pointwise global Galilean invariance
    on the level the kinetic energy for the P-SFS approach.

\subsection{Breaking of Galilean invariance of the SFS stress tensor for non-projector filters} \label{app:GalInv_filter}
We consider the momentum balance for the resolved field $\tbv$ 
with a filtered inertial term  
\be
\label{eq:momentum-ples}
\p_t \tv_i + \p_j(\widetilde{\tv_i\tv_j} + \tP \delta_{ij} + \ttau_{ij}) = 0 \ .
\ee
Under the Galilean transformation given by Equations~\eqref{eq:galilean-transf1}
and \eqref{eq:galilean-transf2}, this equation becomes
\begin{align}
\p_t[\tv_i(\bx-\bu^0t,t)+ u_i^0] + \p_j \left(\extremetilde{(\tv_j+u_j^0)(\tv_i+u_i^0)} + \tP \delta_{ij} 
    +  \ttau_{ij}(\bv+\bu^0 ,\bv+\bu^0 ) \right) =0 \ .
\label{eq:mom-ples-trans}
\end{align}
The subgrid tensor $\ttau_{ij}$, \obs{which, unlike in the main text, 
in this appendix originates from the \emph{filtered} inertial term, i.e., 
$\ttau_{ij}= \widetilde{v_i v_j}-\widetilde{\tv_i \tv_j}$} 
is now not Galilean invariant since
\begin{align}
\label{eq:tau_ng}
\ttau_{ij}(\bv+\bu^0 ,\bv+\bu^0) &
= \extremetilde{(v_i + u_i^0 ) (v_j + u_j^0)} - \extremetilde{(\tv_i + u_i^0 ) (\tv_j + u_j^0)} \nonumber \\ 
&= \widetilde{v_i v_j} + \tv_i u_j^0 + \tv_j u_i^0 + u_i^0 u_j^0 - (\widetilde{\tv_i \tv_j} 
+ \tilde \tv_i u_j^0 + \tilde \tv_j u_i^0 + u_i^0 u_j^0)  \nonumber \\
&=  \widetilde{v_i v_j}-\widetilde{\tv_i \tv_j} + (\tv_j-\tilde \tv_j )u_i^0 + (\tv_i-\tilde \tv_i )u_j^0  \neq \ttau_{ij}(\bv,\bv) \ .
\end{align}
Equation~\eqref{eq:mom-ples-trans} is still {\em globally} Galilean invariant, since
\begin{align}
\p_t \tv_i - u_j^0 \p_j \tv_i & + \p_j(\widetilde{\tv_j\tv_i} 
+ u_j^0 \tv_i +u_i^0 \tv_j 
+ \tilde \tv_i u_j^0 + \tilde \tv_j u_i^0
+ \tP \delta_{ij} +  \ttau_{ij} \nonumber \\ 
&+ (\tv_j-\tilde \tv_j )u_i^0 + (\tv_i-\tilde \tv_i )u_j^0)
= 0 \ ,
\end{align}
where  we recover Equation~\eqref{eq:momentum-ples} because $- u_j^0 \p_j \tv_i$ cancels with $\p_j(u_j^0 \tv_i)$, $\p_j(u_i^0 \tv_j) = 0$
by incompressibility of $\tbv$ and the double-filtered terms cancel out.
For projector filters, we can see directly from Equation~\eqref{eq:tau_ng} that 
$\otau_{ij}$ is Galilean invariant since the terms of the form $\obv - \overline{\obv}$ vanish identically 
because  $G^2=G$. 

\subsection{Global Galilean invariance of the  P-SFS energy balance}
The balance equation for the total resolved energy in the P-SFS approach is
\be
\frac{1}{2}\p_t (\ov_i \ov_i)   +  \p_j \left( \ov_i (\overline{\ov_j \ov_i} + \overline{p}\delta_{ij} + \otau_{ij})\right)   =  -\oPi +  (\p_j \ov_i)(\overline{\ov_j \ov_i}) \ .
\label{eq:ekin-ples}
\ee
Under the Galilean transformation given by Equations~\eqref{eq:galilean-transf1} and \eqref{eq:galilean-transf2}, Equation~\eqref{eq:ekin-ples} becomes
\begin{align}
\frac{1}{2} \p_t[\ov_i(\bx-\bu^0t,t)+ u_i^0]^2 & + \p_j \left( (\ov_i+ u_i^0) \left[\overline{(\ov_j+u_j^0)( \ov_i+u_i^0)} + \overline{p}\delta_{ij} + \otau_{ij}\right] \right)  \nonumber \\ 
& =  -\oPi +  (\p_j (\ov_i+u_i^0))\overline{(\ov_j+u_i^0) )\ov_i+u_i^0)} \ , 
\label{eq:ekin-ples1}
\end{align}
where we used the fact that $\oPi$ and $\tau_{ij}^P$ are Galilean invariant.
We now calculate the remaining terms in Equation~\eqref{eq:ekin-ples1} explicitly. The terms on the LHS are
\begin{align}
\frac{1}{2} \p_t[\ov_i(\bx-\bu^0t,t)+ u_i^0]^2 & = (\ov_i + u_i^0)\p_t \ov_i(\bx-\bu^0t,t) = (\ov_i + u_i^0) (\p_t \ov_i - u_j^0 \p_j \ov_i) \nonumber \\
& = \ov_i \p_t \ov_i + u_i^0 \p_t \ov_i - \ov_i u_j^0 \p_j \ov_i - u_i^0 u_j^0 \p_j \ov_i \ , 
\label{eq:ekin-ples-lhs1}
\end{align}
and
\begin{align}
\p_j \left( (\ov_i+ u_i^0) \overline{(\ov_j+u_j^0)( \ov_i+u_i^0)}\right) & 
= \p_j (\ov_i \overline{\ov_j \ov_i}) + 2u_i^0 u_j^0 \p_j \ov_i + u_j^0 \p_j(\ov_i \ov_i) \nonumber \\ 
& \quad + u_i^0 \ov_j \p_j \ov_i + u_i^0 \p_j\overline{(\ov_i \ov_j)} \ , 
\label{eq:ekin-ples-lhs2}
\end{align}
using incompressibility of $\bv$. The remaining term on the RHS of Equation~\eqref{eq:ekin-ples} is
\begin{align}
(\p_j (\ov_i+u_i^0))\overline{(\ov_j+u_j^0) (\ov_i+u_i^0)}& = (\overline{\ov_j \ov_i} + \ov_i u_j^0 + \ov_j u_i^0 + u_i^0 u_j^0)\p_j \ov_i \ .
\label{eq:ekin-ples-rhs1}
\end{align}
By substitution of the relevant terms with their explicit expressions given in Equations~\eqref{eq:ekin-ples-lhs1}-\eqref{eq:ekin-ples-rhs1},
the kinetic energy budget Equation~\eqref{eq:ekin-ples} becomes
\begin{align}
& \ov_i \p_t \ov_i + u_i^0 \p_t \ov_i - {\ov_i u_j^0 \p_j \ov_i} - {u_i^0 u_j^0 \p_j \ov_i} \nonumber \\ 
& \ \ + \p_j (\ov_i \overline{\ov_j \ov_i}) + {2u_i^0 u_j^0 \p_j \ov_i} + {u_j^0 \p_j(\ov_i \ov_i)} + {u_i^0 \ov_j \p_j \ov_i} + u_i^0 \p_j\overline{(\ov_i \ov_j)} \nonumber \\
& \ \ + \p_j(\ov_i \tau_{ij}^P) + u_i^0 \p_j \tau_{ij}^P + \p_j(\ov_i \overline{p} \delta_{ij}) + u_i^0 \p_j \overline{p} \delta_{ij}\nonumber \\
& \ \ = \overline{\ov_j \ov_i}\p_j \ov_i + {\ov_i u_j^0\p_j \ov_i} + {\ov_j u_i^0\p_j \ov_i} + {u_i^0 u_j^0\p_j \ov_i} - \oPi \ , 
\end{align}
which can be rearranged to
\begin{align}
\label{eq:ples-cancelled}
& \frac{1}{2}\p_t (\ov_i)^2 + \p_j \left( \ov_i (\overline{\ov_j \ov_i} + \overline{p}\delta_{ij} + \otau_{ij})\right) 
 + u_i^0 \left( \p_t \ov_i + \p_j(\overline{\ov_i\ov_j} + \overline{p}\delta_{ij} + \tau_{ij}^P) \right) \nonumber \\
& \ \ = \overline{\ov_j \ov_i} \p_j \ov_i - \oPi \ , 
\end{align}
where we observe that $u_i^0 \left( \p_t \ov_i + \p_j(\overline{\ov_i\ov_j} + \overline{p}\delta_{ij} + \tau_{ij}^P) \right) =0$
from the P-SFS momentum equation.
Hence we recover Equation~\eqref{eq:ekin-ples}
\begin{align}
 \frac{1}{2}\p_t (\ov_i)^2 + \p_j \left( \ov_i (\overline{\ov_j \ov_i} + \overline{p}\delta_{ij} + \otau_{ij})\right) 
 = \overline{\ov_j \ov_i} \p_j \ov_i - \oPi \ , 
\end{align}
and conclude that the P-SFS kinetic energy evolution equation is globally Galilean invariant.\\

\section{Scaling estimates for projector filters} \label{app:scaling}
In this appendix we show that Eyink's proof of Equation~\eqref{eq:interm_scaling} in Ref.~\cite{Eyink05} 
for the F-SFS energy flux extends to the P-SFS 
flux by approximating the sharp Galerkin projector 
with a smooth filter to arbitrary accuracy. 
This statement will be made more precise in the following.
For convenience, we work in $\mathbb{R}^n$. First we establish that Equation~\eqref{eq:interm_scaling} holds
in approximation for the F-SFS formulation and then extend the result to the P-SFS formulation.  
Let $S(\mathbb{R}^n)$ be the Schwartz class of all smooth functions whose derivatives tend to zero 
faster than any power. The elements of $S(\mathbb{R}^n)$ themselves 
also decrease sufficiently fast at infinity. 
Schwartz functions therefore satisfy all the regularity requirements for filters which were 
necessary in the proof of Equation~\eqref{eq:interm_scaling} carried out in Ref.\cite{Eyink05}.
The Schwartz class also has the useful property that 
the Fourier transform maps $S(\mathbb{R}^n)$ to itself (it is an automorphism on $S(\mathbb{R}^n)$). 
In order to find a smooth filter that approximates the sharp projector $\hat G_\Delta$,
it suffices to use a standard result
from functional analysis, namely that $S(\mathbb{R}^n) \subset L^p(\mathbb{R}^n)$ as a dense subspace
for $1 \leqslant p < \infty$. 
Hence if $f \in L^p(\mathbb{R}^n)$ for $1 \leqslant p < \infty$ then we can always find a function 
$f^\eps \in S(\mathbb{R}^n)$ such that 
\be
\|f^\eps-f\|_p < \eps_p \qquad \text{for } 1 \leqslant p < \infty \ , 
\ee 
for any $\eps_p >0$.
For the standard projector $\hat G_\Delta = \theta(k_c-|\bk|)$, where $\theta$ is the Heaviside step function,
it is immediately clear that  $\hat G_\Delta \in L^p(\mathbb{R}^n)$ for $1 \leqslant p \leqslant \infty$, 
and its inverse Fourier transform satisfies $G_\Delta \in L^p(\mathbb{R}^n)$ for $2 \leqslant p \leqslant \infty$.
This implies that we can always find $G_\Delta^\eps \in S(\mathbb{R}^n)$ such that
\be
\|G_\Delta^\eps- G_\Delta\|_p < \eps_p \qquad \text{for } 2 \leqslant p < \infty \ . 
\ee 
Hence we can always approximate the projector filter with a smooth filter which satisfies 
the scaling estimate Equation~\eqref{eq:interm_scaling} for the F-SFS fomulation.
\\

\noindent
The only difference between the P-SFS and the F-SFS formulations is in the 
definition of the SFS stress tensor
\begin{align}
\label{eqapp:ples-decomposed}
\otau(\bv,\bv) &= 
\overline{\bv \otimes \bv} - \overline{\obv \otimes \obv} 
= \overline{\bv \otimes \bv} - \obv \otimes \obv
-(\overline{\obv \otimes \obv} - \obv \otimes \obv) \nonumber \\
& = \underbrace{\overline{\bv \otimes \bv} - \obv \otimes \obv}_\text{F-SFS} 
\underbrace{-(\overline{\obv \otimes \obv} - \overline{\obv} \otimes \overline{\obv})
+(\obv \otimes \obv - \overline{\obv} \otimes \overline{\obv})}_\text{Leonard stress} \ ,
\end{align}
where the Leonard stress has been decomposed into two components, 
the last of which does not vanish in the present case 
since $\overline{(\cdot)}$ here refers to the filtering by $G_\Delta^\eps$ which is not a projector.
According to Ref.~\cite{Eyink05} Equation~(81), the F-SFS component satisfies
\be
\label{eq:fles-estimate}
\|\overline{\bv \otimes \bv} - \obv \otimes \obv\|_p = O(\Delta^{2\zeta_{2p}/p}) \ .
\ee 
The term $\overline{\obv \otimes \obv} - \overline{\obv} \otimes \overline{\obv}$ has also been considered in Ref.~\cite{Eyink05} in connection with the 
infrared locality of the F-SFS stress tensor. In particular, Equation~(99) of 
Ref.~\cite{Eyink05} implies 
\be
\label{eq:leo-estimate}
\|\overline{\obv \otimes \obv} - \overline{\obv} \otimes \overline{\obv}\|_p = O(\Delta^{2\zeta_{2p}/p}) \ .
\ee 
Hence two out of the three terms on the RHS of Equation~\eqref{eqapp:ples-decomposed} have the 
desired scaling properties. In order to obtain the scaling for $\otau(\bv,\bv)$ 
we must consider the $L^p$-norm of the remaining 
term $\obv \otimes \obv - \overline{\obv} \otimes \overline{\obv}$ for the filter $G_\Delta^\eps$
\begin{align}
\label{eqapp:otau-estimate0}
\|\obv \otimes \obv - \overline{\obv} \otimes \overline{\obv}\|_p
&= 
\frac{1}{2}\left(\|(\obv - \overline{\obv}) \otimes(\obv + \overline{\obv}) +
(\obv + \overline{\obv}) \otimes(\obv - \overline{\obv})\|_p\right) \nonumber \\
&\leqslant \|(\obv - \overline{\obv})\|_{2p}\|(\obv + \overline{\obv})\|_{2p}
\nonumber \\
&\leqslant \|G_\Delta^\eps - (G_\Delta^\eps * G_\Delta^\eps)\|_2
          \|G_\Delta^\eps + (G_\Delta^\eps * G_\Delta^\eps)\|_2 \|\bv\|_{r}^2 \ ,
\end{align}
where in the last step we used Young's inequality with $r=2p/(p+1)$
combined with the fact that 
$G_\Delta^\eps \in S(\mathbb{R}^n)$, since this implies that 
$(G_\Delta^\eps * G_\Delta^\eps) \in S(\mathbb{R}^n)$ and 
$G_\Delta^\eps \in L^p(\mathbb{R}^n)$
for $2\leqslant p \leqslant \infty$. The same holds for $(G_\Delta^\eps * G_\Delta^\eps)$. 
The term  
$\|G_\Delta^\eps - (G_\Delta^\eps * G_\Delta^\eps)\|_2 =\|\hat G_\Delta^\eps - (\hat G_\Delta^\eps)^2\|_2$
can now be further estimated
\begin{align}
\label{eqapp:filter-estimate}
\|\hat G_\Delta^\eps - (\hat G_\Delta^\eps)^2\|_2 
&= \|\hat G_\Delta^\eps -\hat G_\Delta - \big((\hat G_\Delta^\eps)^2-\hat G_\Delta\big)\|_2
\leqslant 
\|\hat G_\Delta^\eps -\hat G_\Delta\|_2 +\|(\hat G_\Delta^\eps)^2-\hat G_\Delta\|_2 \nonumber \\
&\leqslant \|\hat G_\Delta^\eps -\hat G_\Delta\|_2 +
\|\hat G_\Delta^\eps-\hat G_\Delta\|_2\|\hat G_\Delta^\eps+\hat G_\Delta\|_\infty \nonumber \\
& \leqslant 
\|\hat G_\Delta^\eps-\hat G_\Delta\|_2\left(1+\|\hat G_\Delta^\eps\|_\infty + \|\hat G_\Delta\|_\infty \right) \ .
\end{align}
where in the third step we used the H\"older inequality and the projector property $\hat G_\Delta^2= \hat G_\Delta$. 
Combining Equations~\eqref{eqapp:otau-estimate0} and \eqref{eqapp:filter-estimate} results in 
\begin{align}
\label{eqapp:otau-estimate1}
\|\obv \otimes \obv - \overline{\obv} \otimes \overline{\obv}\|_p
&\leqslant \|\hat G_\Delta^\eps-\hat G_\Delta\|_2\left(1+\|\hat G_\Delta^\eps\|_\infty + \|\hat G_\Delta\|_\infty \right)  
          \|G_\Delta^\eps + (G_\Delta^\eps * G_\Delta^\eps)\|_2 \|\bv\|_{r}^2 \nonumber \\ 
&\leqslant C\eps \ ,
\end{align}
with $C\equiv \|G_\Delta^\eps + (G_\Delta^\eps * G_\Delta^\eps)\|_2 \|\bv\|_{r}^2
\left(1+\|\hat G_\Delta^\eps\|_\infty + \|\hat G_\Delta\|_\infty \right)$, where we set $\eps \equiv \eps_2$. 
Since $\eps$ can be made arbitrarily small, the P-SFS formulation is expected to satisfy the same 
scaling estimate as the F-SFS formulation, provided the scaling suggested by the bounds in 
Equations~\eqref{eq:fles-estimate} and \eqref{eq:leo-estimate} holds.
While the bounds in Equations~\eqref{eq:fles-estimate} and \eqref{eq:leo-estimate}
formally diverge as $G_\Delta^\eps$ approaches $G_\Delta$, 
we find numerically in Figures~\ref{fig:pi_sq_cbH1} and \ref{fig:pi_sq_cb} that the scaling of these bounds 
remains true when using a sharp Galerkin projector.  
After all, Equations~\eqref{eq:fles-estimate} and \eqref{eq:leo-estimate} are upper bounds and are not necessarily 
tight (sharp) bounds. 
The $L^p$-norm of the SFS stress tensor can (and does) still scale according to the estimate above.

\bibliographystyle{unsrt}
\bibliography{refs,refs2}

\end{document}